\documentclass[fleqn,usenatbib]{mnras}

\usepackage{newtxtext,newtxmath}
\usepackage{graphicx}
\usepackage{bm}
\usepackage{cancel}
\usepackage{epsfig}
\usepackage{dcolumn}
\usepackage{amsmath}
\usepackage{hhline}
\usepackage{enumerate}
\usepackage[utf8]{inputenc}
\usepackage[dvipsnames]{xcolor}
\usepackage{hyperref}
\usepackage{mwe}
\usepackage{widetext}
\usepackage{wasysym}
\usepackage{longtable}
\usepackage{pdflscape}
\usepackage{tikz} 
\usepackage[T1]{fontenc}

\DeclareRobustCommand{\VAN}[3]{#2}
\let\VANthebibliography\thebibliography
\def\thebibliography{\DeclareRobustCommand{\VAN}[3]{##3}\VANthebibliography}

\def \beq  {\begin{equation}}
\def \eeq  {\end{equation}}
\def \ber  {\begin{eqnarray}}
\def \eer  {\end{eqnarray}}

\title{The density of virialized clusters as a probe of dark energy}

\author[Evangelos A. Paraskevas \& Leandros Perivolaropoulos]{
Evangelos A. Paraskevas \orcidB{},$^{1}$\thanks{Contact e-mail: \href{mailto:e.paraskevas@uoi.gr}{e.paraskevas@uoi.gr}}%
Leandros Perivolaropoulos \orcidA{},$^{1}$\thanks{Contact e-mail: \href{mailto:leandros@uoi.gr}{leandros@uoi.gr}}%
\\
$^{1}$Department of Physics, University of Ioannina, GR-45110, Ioannina, Greece
}

\pubyear{2023}

\begin{document}

\interfootnotelinepenalty=10000

\newcommand{\newc}{\newcommand}

\newcommand{\orcidauthorA}{0000-0001-9330-2371} 
\newcommand{\orcidauthorB}{0009-0000-2112-1619} 

\newcommand{\be}{\begin{equation}}
\newcommand{\ee}{\end{equation}}
\newcommand{\ba}{\begin{eqnarray}}
\newcommand{\ea}{\end{eqnarray}}
\newcommand{\bea}{\begin{eqnarray*}}
\newcommand{\eea}{\end{eqnarray*}}
\newc{\D}{\partial}
\newc{\ie}{{\it i.e.} }
\newc{\eg}{{\it e.g.} }
\newc{\etc}{{\it etc.} }
\newc{\etal}{{\it et al.}}
\newc{\lcdm}{$\Lambda$CDM }
\newc{\lcdmnospace}{$\Lambda$CDM}
\newc{\wcdm}{$w$CDM }
\newc{\plcdm}{Planck18/$\Lambda$CDM }
\newc{\plcdmnospace}{Planck18/$\Lambda$CDM}
\newc{\omomnospace}{$\Omega_{0m}$}
\newcommand{\nn}{\nonumber}
\newc{\ra}{\Rightarrow}
\newc{\baodv}{$\frac{D_V}{r_s}$ }
\newc{\baodvnospace}{$\frac{D_V}{r_s}$}
\newc{\baoda}{$\frac{D_A}{r_s}$ } 
\newc{\baodanospace}{$\frac{D_A}{r_s}$}
\newc{\baodh}{$\frac{D_H}{r_s}$ }
\newc{\baodhnospace}{$\frac{D_H}{r_s}$}

\newcommand{\orcidicon}{\includegraphics[width=0.32cm]{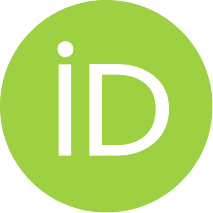}}

\foreach \x in {A, ..., Z}{%
\expandafter\xdef\csname orcid\x\endcsname{\noexpand\href{https://orcid.org/\csname orcidauthor\x\endcsname}{\noexpand\orcidicon}}
}

\label{firstpage}
\pagerange{\pageref{firstpage}--\pageref{lastpage}}
\maketitle

\begin{abstract}
We use the spherical collapse model to demonstrate that the observable average density of virialized clusters depends on the properties of dark energy along with the properties of gravity on cluster scales and can therefore be used as a probe of these properties. As an application of this approach we derive the predicted virialized densities and radii of cluster mass structures for a wide range of values of the cosmological constant (including negative values) as a function of the turnaround redshift. For the value of $\Omega_{\Lambda,0}=-0.7$ (with $\Omega_{m,0}=0.3$) preferred by $\Lambda$ sign-switching models ($\Lambda_s\text{CDM}$) proposed for the resolution of the Hubble and $S_8$ tensions,  we find an amplification of the density of virialized clusters which can be as large as $80\%$ compared to \plcdm for a turnaround redshift $z_{\text{max}} \gtrsim 2$. Such an amplification may lead to more efficient early galaxy formation in this class of models in accordance with the recent findings of JWST.
\end{abstract}




\section{Introduction}
\label{intro}

The standard \lcdm model \citep{Carroll:1991mt,Peebles:2002gy,Bull:2015stt,Dodelson:2003ft} is based on the assumption that the energy density of the universe is currently dominated by two components: $70\%$ dark energy whose role is played by a positive cosmological constant with equation of state $w=-1$ and about $30 \%$ of a dark matter + baryons fluid with equation of state $w=0$. Despite of a wide range of successful predictions \citep{Planck:2018vyg,KiDS:2020ghu} in a wide range of cosmological data, two parameters of the model, the Hubble constant $H_0$ and the growth of perturbations strength parameter $S_8$, have best fit values that are inconsistent among different cosmological probes and in particular between Cosmic Microwave Background (CMB) measurements and local universe measurements. The level of these {\it tensions}, the Hubble tension \citep{Verde:2019ivm,DiValentino:2021izs,Perivolaropoulos:2021jda,Abdalla:2022yfr} and the growth tension \citep{DiValentino:2020vvd,Perivolaropoulos:2021jda,Abdalla:2022yfr}, is in the range of $3-5\sigma$. The persistence of these (and other less discussed \citep{Perivolaropoulos:2021jda})  tensions during the past 5-10 years has lead to the possible  anticipation that new physics may be the their origin \citep{Perivolaropoulos:2021jda,Abdalla:2022yfr,Hu:2023jqc}. 

A wide range of theoretical models have been constructed in an effort to extend the standard \lcdm model introducing new degrees of freedom which could eliminate these tensions and  lead to a consistency of the local late time probes with the early time cosmological probes. These models may be classified in three broad classes according the cosmic time when they introduce new physics degrees of freedom 
\begin{itemize}
    \item {\bf Early time models:}  These models introduce new physics degrees of freedom just before recombination (redshift $z\gtrsim 1100$) decreasing the sound horizon distance calibrator scale at recombination. Examples of these theoretical frameworks encompass models such as Early Dark Energy, outlined in \cite{Karwal:2016vyq,Poulin:2018cxd,Poulin:2018dzj,Agrawal:2019lmo,Kamionkowski:2022pkx,Odintsov:2023cli}, in addition to the paradigm of New Early Dark Energy, explicated in \cite{Niedermann:2019olb,Niedermann:2023ssr}.
    \item {\bf Intermediate/late time models:} These models introduce new physics degrees of freedom at intermediate/late times ($z\simeq 0.1-3$) deforming the functional shape of the Hubble expansion rate $H(z)$. Examples of these theoretical frameworks encompass models such as Interacting Dark Energy \citep{Pan:2019gop,DiValentino:2019jae,DiValentino:2019ffd,Kumar:2021eev}, Phantom Crossing Dark Energy \citep{DiValentino:2020naf,Alestas:2020mvb,Alestas:2020zol,Gangopadhyay:2022bsh,Gangopadhyay:2023nli}, Vacuum Metamorphosis (VM) model \citep{DiValentino:2020kha,DiValentino:2021rjj}, Clustering Dark Energy \cite{Batista:2021uhb,Heisenberg:2022lob}, or the $\Lambda_s \text{CDM}$ model \citep{Akarsu:2019hmw,Akarsu:2021fol,Akarsu:2022typ,Adil:2023ara,Akarsu:2023mfb}. 
    \item {\bf Ultralate time models:} These models introduce new physics degrees of freedom at ultralate times ($z\lesssim 0.01$) affecting the astrophysics and/or gravitational properties of distance scale standard candle and standard ruler calibrators \citep{Marra:2021fvf,Alestas:2020zol,Alestas:2021nmi,Alestas:2021luu,Perivolaropoulos:2021bds,Perivolaropoulos:2021jda}. 
\end{itemize}
Since, the shape of the Hubble free expansion rate $E(z)\equiv H(z)/H_0$ is very well constrained by SnIa standard candles and Baryon Acoustic Oscillations (BAO) standard ruler data at redshifts $0.01<z<3$, the most successful representative models of each one of the above classes involve some kind of abrupt event (transition) which leaves practically unaffected the model consistency with the data before and after its occurrence. Such {\it transition} based models include the 'New Early Dark Energy' model \citep{Niedermann:2019olb,Niedermann:2020dwg}  which involves a first order dark energy phase transition occurring just before recombination \citep{Niedermann:2021vgd,Niedermann:2021ijp}, the $\Lambda_s\text{CDM}$ model \citep{Akarsu:2021fol,Akarsu:2022typ,Akarsu:2023mfb} which involves a sign switch of the cosmological constant from a negative value to a positive value at around $z\simeq 2$ and the gravitational transition hypothesis which involves a sudden change of the strength of gravity by a few percent either locally or globally in the Universe during the past $150\,\text{Myrs}$ ($z\lesssim 0.01$) \citep{Marra:2021fvf}.

There are indications of the possibility of negative Dark Energy (DE) densities at higher redshifts, as observed in Lyman-$\alpha$ (2014) and Pantheon+ data \citep{Aubourg:2014yra,refId0,Mortsell:2018mfj,duMasdesBourboux:2020pck,Colgain:2022rxy,Malekjani:2023dky} (note that Lyman-$\alpha$ (2020) BAO data show less evidence for this trend\citep{Colgain:2023bge}). An interesting cosmological probe that can be used to simulatenously probe gravitational and dark energy degrees of freedom involves the nonlinear clustering of cosmological structures including clusters and galaxies. An efficient analytical method to study the predicted properties of nonlinear virialized structures in the context of cosmological models is the {\it spherical collapse model} \citep{Gunn:1972sv}. 

The assumption of spherical symmetry allows the analytical investigation of the evolution of a spherically symmetric overdensity in an expanding universe beyond the linear approximation and up to the virialization of the collapsing structure where the system has relaxed to a final equlibrated nonlinear configuration where the virial theorem is applicable. 
In the process of collapse, it is postulated that if the density of dark matter within a certain closed region of the cosmos surpasses the background matter density, then a collapse ensues \citep{Press:1973iz,Bond:1990iw,Cooray:2002dia,Asgari:2023mej,Dodelson:2003ft}. In the context of spherical collapse, the initially overdense closed region expands in tandem with the background expansion. However, this expansion eventually ceases due to attractive gravitational forces, leading to a turnaround. After this turnaround, the closed region begins to collapse. A purely general relativistic spherical collapse typically results in a singularity as the final state \citep{Penrose:1964wq}. Nevertheless, in astrophysics, it is generally assumed that the collapsing fluid virializes long before a singularity forms. The virialized final state of the collapse signifies the formation of structure. In this regard, the spherical collapse is a semi-relativistic process where a semi-Newtonian paradigm is employed to interpret the final phase of the collapse. 

The spherical collapse process involves two critical points in time: the time of turnaround of a given spherical shell when the shell of matter stops expansion and stands still before starting to collapse and the time of virialization when the system equilibrates under the validity of the virial theorem. The radii and densities of the spherical collapsing structures at these two critical times constitute observables which are sensitive to the properties of gravity and dark energy as manisfested during the collapse process. 

In a series of papers, Pavlidou and collaborators \citep{Pavlidou:2004vq,Pavlidou:2013zha,Pavlidou:2014aia,Tanoglidis:2016lrj,Korkidis:2019nzk,Pavlidou:2020afx} have demonstrated that the observed turnaround radii and densities of collapsing clusters can lead to constraints on the value of the cosmological constant which are independent from the standard cosmological distance probes.  In the present analysis we demonstrate that the corresponding virialized radii and densities can play a similar role as observational probes that can lead to constraints on the properties of dark energy. The observational identification of the properties of virialized structures is distinct and perhaps easier to measure than the observational identification of structures that are at the turnaround stage. 

The structure of this paper is the following: In the next section we present the main steps and equations used for the  calculation of the turnaround and virialized radii and densities of cosmological structures as functions of the turnaround redshift in models with dark energy. In section 3 we apply these equations to models based on a positive or negative cosmological constant and find the ratios of the virialized radii and densities relative to \plcdm  as  functions of the turnaround redshift. Finally in section 4 we conclude, summarize our results and discuss possible future extensions of our analysis.

\section{The spherical collapse model in the presence of dark energy}
\label{sec2}

\subsection{Spherical Collapse}
The spherical collapse model in the presence of dark energy has been discussed previously in \cite{Lokas:2000cn,Mota:2004pa,Basilakos:2009mz,Creminelli:2009mu,Tanoglidis:2014lea,Tanoglidis:2016lrj,Pavlidou:2004vq,Pavlidou:2013zha,Pavlidou:2014aia,Korkidis:2019nzk,Pavlidou:2020afx} either up to the turnaround phase or from the turnaround phase up to the virialization phase. Here we combine the two phases in a unified approach leading to predictions for the cosmological observable virialized density \citep{Pace:2017qxv}.

Let's assign $\rho$ as the symbol to represent the energy density of the spherical overdensity, while using $\tilde{\rho}$ to denote the energy density of the background universe. Let us denote the overdensity at the  cosmic epoch corresponding to a background scale factor  $a$ as
\begin{equation}
\delta(a)=\frac{\rho_{m}(a)-\tilde{\rho}_{m}(a)}{\tilde{\rho}_{m}(a)}
\end{equation}
 With these definitions, the homogeneous spherical overdensity at the cosmic time $t$, can be expressed as
\begin{equation}\label{eq08}
\rho_{m}\left(a\right)=\tilde{\rho}_{m}\left(a\right)[1+\delta(a)]=\tilde{\rho}_{m,0}a^{-3}[1+\delta(a)]
\end{equation} 
In this context, we introduce the symbol $\tilde{\rho}_{m,0}$ to represent the matter density of the background universe in the current epoch. By incorporating the normalization, as discussed in the paper \cite{Pavlidou:2004vq}, 

\begin{equation}\label{eq001}
R^{-3}(a)=a^{-3}[1+\delta(a)]
\end{equation}
the homogenous spherical overdensity (see Equation (\ref{eq08})) is written as
\begin{equation}
\rho_{m}(a)=\tilde{\rho}_{m,0}R^{-3}(a)
\end{equation}
We define the instance in time at which the overdense region reaches its maximal extent as $t=t_{\text{max}}$ (turnaround). Correspondingly, the background scale factor at this specific time, $t_{\text{max}}$, is denoted as $a(t_{\text{max}})\equiv a_{\text{max}}$. Furthermore, the local scale factor $R$ of the spherical overdensity, when the scale factor reaches $a_{\text{max}}$, is denoted as $R(a_{\text{max}})\equiv R_{\text{max}}$. We also denote the overdensity at turnaround time as $\delta(a_{\text{max}})\equiv\delta_{\text{max}}$. Hence, the density of the homogeneous spherical overdensity at the moment of turnaround is
\begin{equation}
\rho_{m,\text{max}}\equiv\rho_{m}\left(a_{\text{max}}\right)=\tilde{\rho}_{m,0}R_{\text{max}}^{-3}
\end{equation}
From Equation (\ref{eq001}), we get the overdensity at the moment of turnaround as \citep{Pavlidou:2013zha,Pavlidou:2014aia,Tanoglidis:2016lrj}
\begin{equation}\label{eq009}
\delta_{\text{max}}=\left(\frac{a_{\text{max}}}{R_{\text{max}}}\right)^3-1
\end{equation}
Given that \( R \) grows at a slower rate than \( a \). This implies that the matter density within the overdensity, \( \tilde{\rho}_{m,0}R^{-3} \), exceeds the background matter density, \( \tilde{\rho}_{m,0}a^{-3} \). In the forthcoming sections, we use the notation $R(t)$ to represent the local scale factor within the spherical overdensity, wherein, $R_{p}$ the physical radius  of a local overdensity is  $R_{p}\equiv R_{p}(t)=\chi_{0}R(t)$, where $\chi_{0}$ is the corresponding comoving radius.

The dynamics of both $R(t)$ and $a(t)$ are determined by two Friedman equations which include spatial curvature matter density and cosmological constant. The corresponding Friedmann equation for the background universe is written as \citep{Dodelson:2003ft}

\be\label{afried}
    \left(\frac{\dot{a}}{a}\right)^2=\frac{8\pi G}{3} \left(\tilde{\rho}_{m,0}a^{-3}+\rho_{\Lambda}+\frac{\tilde{\rho}_{\text{crit},0}-\tilde{\rho}_{m,0}-\rho_{\Lambda}}{a^2}\right)
\ee
The energy density attributed to the cosmological constant is denoted by $\rho_{\Lambda}\equiv \frac{\Lambda}{8\pi G}$. Furthermore, the critical density of the background at the current epoch is \begin{equation}\tilde{\rho}_{\text{crit},0}\equiv \frac{3 H_{0}^2}{8\pi G}\end{equation}

We also introduce a term $\omega$, defined as the ratio of the energy density due to the cosmological constant and the present day background matter density, i.e. \begin{equation}
\omega\equiv \frac{\rho_{\Lambda}}{\tilde{\rho}_{m,0}}=\frac{\Omega_{\Lambda,0}}{\Omega_{m,0}}
\label{omegadef}
\end{equation}
 
The parameter associated with the spatial curvature of the cosmological background  are further described 
\begin{equation}\xi\equiv\frac{\tilde{\rho}_{\text{crit},0}-\tilde{\rho}_{m,0}-\rho_{\Lambda}}{\tilde{\rho}_{m,0}}=\frac{1}{\Omega_{m,0}}-1-\omega 
\label{xidef}
\end{equation}
Considering the aforementioned details, the Friedman equation is expressed as in Eq.~(\ref{afried}):
\begin{equation}\label{afried1}
    \left(\frac{\dot{a}}{a}\right)^2=\frac{8\pi G}{3} \tilde{\rho}_{m,0}\left(a^{-3}+\omega+\frac{\xi}{a^2}\right)
\end{equation}
Concurrently, we utilize the notation $\rho_{m,s}$ to denote the energy overdensity at an arbitrary reference cosmic time $t_{s}$. The critical density of the spherical region, $\rho_{\text{crit},s}$, at the moment where $t=t_{s}$,  is expressed as 
\begin{equation}
\rho_{\text{crit},s}\equiv\frac{3}{8\pi G}\left(\frac{\dot{R}}{R}\bigg{|}_{t=t_{s}}\right)^2
\end{equation}

\textbf{Proof of Equation (\ref{eq001}): A Brief Note}. Prior to delving into the detailed analysis, it is pertinent to note that all essential quantities required for the elucidation of Equation (\ref{eq001}) have been established. We introduce an initial cosmic time, represented as \( t_{s} \), during which the scale factors coincide, that is,
\begin{equation}\label{initcond}
a(t_{s}) = R(t_{s}).
\end{equation}
Furthermore, we postulate that the velocities at that moment are
\begin{equation}
\dot{R}(t_{s}) = \dot{a}(t_{s}),
\label{dotRs}
\end{equation}
 The parameters associated with the spatial curvatures of both the
cosmological background and the spherically over-dense regions at the initial moment, $t_{s}$, are
 \begin{equation}
     \tilde{\xi}\equiv\frac{\tilde{\rho}_{\text{crit},s}-\tilde{\rho}_{m,s}-\rho_{\Lambda}}{\tilde{\rho}_{m,s}}
 \end{equation}
 
 \begin{equation}
\tilde{\kappa}\equiv \frac{\rho_{m,s}+\rho_{\Lambda}-\rho_{\text{crit},s}}{\tilde{\rho}_{m,s}}
\end{equation} 
 Moreover, drawing from Equation (\ref{dotRs}), one can deduce that the critical densities at the specified time \( t_{s} \) for both scenarios are coinciding as \( \rho_{\text{crit},s} = \tilde{\rho}_{\text{crit},s} \). To complete our set of assumptions, it is essential to note that the curvatures at this time are necessitated as
\begin{equation}\label{curv}
     -\tilde{\kappa} =\tilde{\xi} \implies \rho_{m,s}=\tilde{\rho}_{m,s}
\end{equation}
Provided that the general solution of the following differential equation
\begin{equation}
    \dot{\rho}_{m}+3\frac{\dot{R}}{R}\rho_{m}=0
\end{equation}
is written as 
\begin{equation}\label{solrho}
    \rho_{m}(t)=\frac{c}{R^{3}(t)}
\end{equation}
Then, the initial condition (\ref{initcond}) and Equation (\ref{curv}) imply that 
\begin{equation}\label{const}
    c=\tilde{\rho}_{m,s}R_{s}^3=\tilde{\rho}_{m,0}\frac{R_{s}^3}{a_{s}^3}=\tilde{\rho}_{m,0}
\end{equation}
Drawing upon the combined results from Equations (\ref{solrho}) and (\ref{const}), we obtain:
\begin{equation}
\rho_{m}(a) = \tilde{\rho}_{m,0} R^{-3}(a),
\end{equation}
In the concluding step of the proof, we have 
\begin{equation}
\rho_{m}(a) = \tilde{\rho}_{m}(a) [1 + \delta(a)]
\end{equation}
From which it naturally follows that 
\begin{equation}
R^{-3}(a) = a^{-3} [1 + \delta(a)]. 
\end{equation}

The Friedmann equation corresponding to the local overdensity scale factor \( R(t) \), considering the previously delineated initial conditions, can be deduced as:
\begin{equation}\label{rescale2}
    \left(\frac{\dot{R}}{R}\right)^2=\frac{8\pi G}{3}\tilde{\rho}_{m,s} \left(R^{-3}+\frac{\rho_{\Lambda}}{\tilde{\rho}_{m,s}}-\frac{\tilde{\kappa}}{R^2}\right)
\end{equation}
By considering the relation \(\rho_{m,s} = \tilde{\rho}_{m,s} = \tilde{\rho}_{m,0}a_{s}^{-3}\) and implementing the rescaling transformations \(a_{s} R \rightarrow R\) and \(\frac{\tilde{\kappa}}{a_{s}} \rightarrow \kappa\), Equation~\eqref{rescale2} can be expressed as \citep{Eke:1996ds,Pavlidou:2004vq,Tanoglidis:2016lrj,Pavlidou:2020afx}:
\begin{equation}\label{rfried1}
    \left(\frac{\dot{R}}{R}\right)^2=\frac{8\pi G}{3}\tilde{\rho}_{m,0} \left(R^{-3}+\omega-\frac{\kappa}{R^2}\right)
\end{equation}

When, we divide the preceding two equations (\ref{afried1}),(\ref{rfried1}) part by part, we arrive at the following result \citep{Eke:1996ds,Pavlidou:2004vq,Pavlidou:2013zha,Pavlidou:2014aia,Tanoglidis:2016lrj}:
\begin{equation}\label{dRda}
  \left(\frac{dR}{da}\right)^2 =\frac{R^2}{a^2}\frac{R^{-3}+\omega-\frac{\kappa}{R^2}}{a^{-3}+\omega+\frac{\xi}{a^2}}=\frac{a}{R}\frac{1+\omega R^3-\kappa R}{1+\omega a^3+\xi a}.
\end{equation}
Hence, Equation (\ref{dRda}) implies that
\begin{equation} \label{drda8}
     \frac{dR}{da} =\pm \sqrt{\frac{a}{R}\frac{1+\omega R^3-\kappa R}{1+\omega a^3+\xi a}}
\end{equation}
The overdensity prior to turnaround, is associated with the positive branch of Equation (\ref{drda8}), denoting an expansion phase. Conversely, the negative branch corresponds to a contraction phase, indicative of a collapse. Equation \eqref{dRda} is considered valid up to the turnaround moment; beyond this, shell crossing may intervene. The negative branch, indicative of a collapse phase, is omitted from this analysis.
\subsection{Turnaround stage}

Equation (\ref{drda8}) determines the evolution of the local scale factor $R(t)$ for the spherical overdensity, as a function of the background scale factor $a(t)$. Because of the increased positive spatial curvature $\kappa$ within the spherical overdensity region, its radius $R_{p}$ will undergo a decelerated expansion. It will ultimately initiate a turnaround, and then it will collapse and virialize into a stable, stationary bound state (Section \ref{sec2.3}). At turnaround time $t_{\text{max}}$, the radius $R_{p}$ attains a maximum value denoted as $R_{p,\text{max}}$. For the initial expansion phase of the overdensity (before turnaround) we choose the positive branch of (\ref{drda8}), to obtain \citep{Tanoglidis:2016lrj}, 
\begin{equation} \label{drda2}
     \frac{dR}{da} =\sqrt{\frac{a}{R}\frac{1+\omega R^3-\kappa R}{1+\omega a^3+\xi a}}
\end{equation}
The spherical overdensity will turn around, when the local scale factor reaches its maximum $R_{\text{max}}$. Thus $\kappa$ may be obtained in terms of $R_{\text{max}}$ using Equation (\ref{drda2}) and the condition \begin{equation}\frac{d R}{da}\bigg{|}_{a=a_{\text{max}}}=0\end{equation}
This leads to 
\begin{equation}\label{eq011}
1+\omega R_{\text{max}}^3-\kappa R_{\text{max}}=0
\end{equation}
The solution of Equation (\ref{eq011}), yields the value of the parameter $\kappa$ as a function of $\omega$ and $R_{\text{max}}$:
\begin{equation}\label{kapparmax}
\kappa= \frac{\omega R^3_{\text{max}}+1}{R_{\text{max}}}
\end{equation}

By separating variables and integrating Equation (\ref{drda2}), we get 
\begin{equation}\label{drda3}
\int_{0}^{R}dR'\sqrt{\frac{R'}{1+\omega R'^3-\kappa R'}}=\int_{0}^{a}da'\sqrt{\frac{a'}{1+\omega a'^3+\xi a'}}
\end{equation}

We now set $u=\frac{R'}{R_{\text{max}}}$ in the left-hand side of equation (\ref{drda3}), $y=\frac{a'}{a_{\text{max}}}$ in the right-hand side and use (\ref{kapparmax}) in the form $\kappa R_{\text{max}}^{-2}=\omega+R_{\text{max}}^{-3}$.  Using also (\ref{eq009}) we find
\begin{equation}\label{drda6}
\int_{0}^{1}\sqrt{\frac{1}{1-u}\frac{u}{\frac{1+\delta_{\text{max}}}{\omega a_{\text{max}}^{3}}-u(u+1)}}du =\int_{0}^{1} \sqrt{\frac{\omega y }{\omega y^{3}+\xi y a_{\text{max}}^{-2}+a_{\text{max}}^{-3}}}dy
\end{equation}
We have integrated up to the turnaround radius $R_{p,\text{max}}$ which is equivalent to setting the integral boundaries as $y=1$ and $u=1$\footnote{ A detailed discussion can be found in Appendix \ref{AppA}.}. The numerical solution of Equation (\ref{drda6}) leads to the functions $\delta_{\text{max}}(a_{\text{max}},\Omega_{\Lambda,0},\Omega_{m,0})$ for given values of the parameters $\omega$ and $\xi$ as obtaned from Equations (\ref{omegadef}) and (\ref{xidef}).

Furthermore, if we were to integrate the positive branch of Equation (\ref{drda8}) for an arbitrary value of $R(a)<R_{\text{max}}$, then by proceeding with numerical integration until $y=\frac{a}{a_{\text{max}}}$ and $u=\frac{R(a)}{R_{\text{max}}}$ are reached, we could derive  the local scale factor $R\equiv R(a)$ in terms of the background scale factor $a$.

\subsection{Virialization stage}\label{sec2.3}

For achieving a stable state, we posit that interactions between mass shells cause a shift from a purely general relativistic spherical collapse to a semi-relativistic process. To interpret the final stage of this collapse, we employ a semi-Newtonian framework. Transitioning to the Newtonian limit of Einstein's field equations, while considering the equation of state as $p=w \rho c^2$, leads to a Poisson equation that incorporates pressure as a source of gravitational field:

\begin{equation}\label{Poiss}
\nabla^2 \Phi = 4\pi G \rho(1+3 w)
\end{equation}
Assuming a theoretical framework in which \textit{a homogeneous energy density is distributed spherically within a radius} $R$ and the gravitational potential is $\Phi\equiv\Phi(r)$, we apply the general solution 
\begin{equation}\label{gensol}
\Phi(r)=2 \pi G \rho (1+3w)\frac{r^2}{3}-\frac{C}{r}+D\end{equation}
for both regions $r\leq R$ and $r\geq R$, while admitting that $\rho(r>R)=0$. The following boundary conditions are also applied (see Appendix \ref{AppB}):

\begin{enumerate}
    \item[]\begin{equation}
        \Phi_{<}(r=0)<\infty
    \end{equation}
 \item[]\begin{equation}
        \Phi_{>}(r\to \infty)=0
    \end{equation}
    \item[] \begin{equation}
    \Phi_{>}(R)=\Phi_{<}(R)
    \end{equation}
    \item[]  \begin{equation}
        \frac{d\Phi_{>}}{dr}\bigg{|}_{r=R}=\frac{d\Phi_{<}}{dr}\bigg{|}_{r=R}
    \end{equation}
\end{enumerate}

Subsequently, the solution to the Poisson equation, Equation (\ref{Poiss}), suggests that the gravitational potential for dark matter energy density confined in a sphere of radius $R$ is $\rho_{\text{DM}}$ ($w=0$), is given by

\begin{equation}
\Phi_{\text{DM}}(r)=\begin{cases}
    -2 \pi G \rho_{\text{DM}} (R^2-\frac{r^2}{3})\quad r\leq R\\
-4\pi G \rho_{\text{DM}} \frac{R^3}{3r}=-\frac{GM}{r}\quad r\geq R
    \end{cases}
\end{equation}

In the second framework, in the case of \textit{homogeneity in the energy density of dark energy throughout the universe}, we choose the boundary condition $\Phi(r= 0)=0$. Within this context, the solution to the Poisson equation (\ref{gensol}) provides the gravitational potential 
\begin{equation}
   \Phi_{\text{DE}}= 2 \pi G \rho_{\text{DE}} (1+3w)\frac{r^2}{3}
\end{equation}

Drawing from the works of \cite{Lahav:1991wc,Iliev:2001he,Maor:2005hq,Saha:2023zos}, we describe the gravitational potential of a dual-component system consisting of dark energy and dark matter, where only the dark matter component undergoes virialization. The homogeneity of dark energy density, which seamlessly permeates from the external surroundings into the over-dense regions, indicates that dark energy does not coalesce or virialize with dark matter. In such a scenario, the gravitational potential energy can be expressed as

\begin{equation} \label{pote1}
U_{\text{system}}=\frac{1}{2}\int_{V} \rho_{\text{DM}} \Phi_{\text{DM}} \, dV+\int_{V} \rho_{\text{DM}} \Phi_{\text{DE}} \, dV
\end{equation}

Considering a homogeneous spherical distribution of dark matter with a radius $R_{p}$ and mass $M$, the energy density is given by 
\begin{equation}\rho_{\text{DM}}=\frac{M}{\frac{4}{3}\pi R_{p}^{3}}\end{equation} 
In relation to their values at the time of turnaround, and with \begin{equation}
\epsilon\equiv\frac{\rho_{\text{DE}}(t_{\text{max}})}{\rho_{\text{DM}}(t_{\text{max}})}
\label{epsdef}
\end{equation} 
representing the ratio of dark energy to dark matter energy densities at that specific point, we can express the ratio of energy densities as

\begin{equation}
\frac{\rho_{\text{DE}}(t)}{\rho_{\text{DM}}(t)}=\epsilon\left(\frac{a(t)}{a(t_{\text{max}})}\right)^{-3(1+w)}\left(\frac{R_{p}(t)}{R_{p}(t_{\text{max}})}\right)^3
\end{equation} 
For a homogeneous spherical dark matter distribution of a radius $R_p$ and mass $M$, we get (see Appendix \ref{AppB})

\begin{equation}
    \frac{1}{2}\int_{V} \rho_{\text{DM}}\Phi_{\text{DM}} \, dV=-\frac{3}{5}\frac{GM^2}{R_{p}}
\end{equation}
The integral over the volume of $\rho_{\text{DM}}\Phi_{\text{DE}}$, can then be written as (see Appendix \ref{AppB}): 
\begin{equation}
\int_{V} \rho_{\text{DM}} \Phi_{\text{DE}}\, dV=
\end{equation} 
\begin{equation*}
    =-\frac{3}{5}\frac{GM^2}{R_{p}}\bigg{[}-\frac{1}{2} (1+3w)\epsilon \left(\frac{a}{a_{\text{max}}}\right)^{-3(1+w)}\left(\frac{R_{p}}{R_{p,\text{max}}}\right)^3\bigg{]}
\end{equation*}
Thus, the potential energy (\ref{pote1}) of the system can be expressed as:
\begin{equation} \label{pote2}
U_{\text{system}}=-\frac{3}{5}\frac{GM^2}{R_{p}}\left[1+\theta \left(\frac{a}{a_{\text{max}}}\right)^{-3(1+w)}\left(\frac{R_{p}}{R_{p,\text{max}}}\right)^3\right]
\end{equation}
where we have used the notation for
\begin{equation}
\theta\equiv-\frac{1}{2}(1+3w)\epsilon
\end{equation}
The conservation of energy for the system, denoted as $E$, implies the following equation
\begin{equation}\label{Energy}
E=(K_{\text{system}}+U_{\text{system}})\big{|}_{R_{p}=R_{p,\text{vir}}}=U_{\text{system}}\big{|}_{R_{p}=R_{p,\text{max}}}
\end{equation}
If we assume that the interactions between mass shells instigate a transition from a fully general relativistic spherical collapse to a semi-relativistic process, we can subsequently bifurcate the general relativistic process. This allows us to attach our current derivation with the previously discussed spherical collapse model. According to the \textit{Virial theorem}, the kinetic energy after equilibration and virialization is connected with the potential energy as $K=\frac{R_{p}}{2}\frac{dU}{dR_{p}}$. Thus, using eq. (\ref{Energy}), the virialization condition becomes 
\begin{equation}\label{eq017}
    \bigg(\frac{R_{p}}{2}\frac{dU_{\text{system}}}{dR_{p}}+U_{\text{system}}\bigg)\bigg|_{R_{p}=R_{p,\text{vir}}}=U_{\text{system}}\big{|}_{R_{p}=R_{p,\text{max}}}
\end{equation}
Using Equations \eqref{pote2} and (\ref{eq017}) can be recast the viralization condition takes the form \footnote{A detailed derivation of these steps can be found in Appendix \ref{AppB}.} \citep{Lahav:1991wc,Iliev:2001he,Battye:2003bm,Maor:2005hq,Saha:2023zos} (see also \cite{Wang:1998gt,Weinberg:2002rd,Basilakos:2003bi,Horellou:2005qc})
\begin{equation}\label{eq022} 
    4 \theta \eta^3 \left(\frac{a_{\text{vir}}}{a_{\text{max}}}\right)^{-3(1+w)}-2 (1+\theta)\eta +1=0
\end{equation}
 where we have defined
 \begin{equation}
\eta\equiv \frac{R_{\text{vir}}}{R_{\text{max}}}
\end{equation}
 Due to the inherent complexity of the solution and the fact that we anticipate that the parameter $\theta$ would exhibit a small magnitude for $z>1$, which can be attributed to the comparatively lower density of dark energy in relation to matter during the early stages of cosmic evolution, it is standard practice to employ the following series expansion in terms of $\theta$. For both positive and negative values of $\theta$, Equation (\ref{eq022}) leads to the following  result:

\begin{equation}\label{eq024}
   \eta=\frac{1}{2} + \frac{1}{4} \theta \left(-2 + \alpha\right) + \frac{1}{8} \theta^2 \left(-2 + \alpha\right) \left(-2 + 3 \alpha\right) + 
\end{equation}
\begin{equation*}
    +\frac{1}{8} \theta^3 \left(-2 + \alpha\right) \left(2 - 9 \alpha + 6 \alpha^2\right) + 
\end{equation*}

\begin{equation*}
    +\frac{1}{32} \theta^4 \left(-2 + \alpha\right) \{-8 + \alpha [76 + 5 \alpha \left(-26 + 11 \alpha\right)]\}+...
\end{equation*}
where
\begin{equation}
\alpha\equiv\left(\frac{a_{\text{vir}}}{a_{\text{max}}}\right)^{-3(1+w)}
\end{equation} 

For a thorough elucidation of the derivations and additional context pertaining to these calculations, we kindly direct the reader to consult Appendix \ref{AppB}.

\section{Observational Predictions and Results}

The effects of repulsive gravity induced by dark energy and/or cosmological constant are expected to lead to a reduction of the density of structures both at the turnaround and at the virialization stages. This reduction however is not fully consistent with the recent findings of the James Webb Space Telescope \citep{Menci:2022wia,Biagetti:2022ode,Wang:2022jvx,Forconi:2023izg} indicating a potential for more efficient early galaxy formation at higher redshifts. This discovery along with other cosmological data has lead to the investigation of model with negative dark energy density at early times (eg negative cosmological constant) which would enhance rather than decrease the efficiency of structure formation at early times. For example $\Lambda$ sign-switching cosmologies ($\Lambda_s\text{CDM}$ \citep{Akarsu:2021fol,Akarsu:2022typ,Adil:2023ara}) assume the presence of a negative cosmological constant $\Lambda$ at early times ($z>2$) which switches sign at late times. This class of models appears to have several attractive features as it also has the potential to resolve the Hubble and growth tensions. Thus, this section we quantitatively estimate the increased efficiency of models that involve a negative cosmological constant to efficiently boost nonlinear structure formation and produce structures with turnaround and virialized densities that are significantly larger than \plcdm.  

We thus assume a cosmological constant of arbitrary sign and magnitude and a fixed mass $M$ of the overdensity. Thus the density of the collapsing structures may be denoted as 
\begin{equation}\label{eqrho}
\rho_{m,\text{max}}=\frac{M}{\frac{4}{3}\pi R_{p,\text{max}}^{3}}
\end{equation} 
which represents the constant density within the sphere at the moment of the turnaround. By considering equations (\ref{eq08}) and (\ref{eqrho}), we get
\begin{equation}\label{rmax}
    R_{p,\text{max}}=\left(\frac{3 M}{4 \pi (1+\delta_{\text{max}})\tilde{\rho}_{m,0}}\right)^{\frac{1}{3}}a_{\text{max}}
\end{equation}

Additionally, we establish the notation $M_{c}\equiv 10^{13}M_{\odot}$ to denote a mass that signifies the lower threshold of the galaxy cluster mass range. The corresponding length scale, $R_{c}$  is defined as 
\begin{equation}\label{rc}
R_{c}\equiv\left(\frac{3 M_{c}}{4 \pi \tilde{\rho}_{\text{crit},0}}\right)^{\frac{1}{3}}
\end{equation}
while the ratio of equations, (\ref{rmax}) and (\ref{rc}), gives
\begin{equation}\label{ratiorc}
    \frac{R_{p,\text{max}}}{R_{c}}=\left(\frac{M}{M_{c}\left(1+\delta_{\text{max}}\right)\Omega_{m,0}}\right)^{\frac{1}{3}}a_{\text{max}}
\end{equation}
This expression allows the quantification of the turnaround radius relative to the aforementioned characteristic scale.

In the specific case of a cosmological constant where $w=-1$, we have $\theta=\epsilon$, and consequently, $\alpha=1$. Thus Equation (\ref{eq024}) takes
\begin{equation}\label{rvirrmax}
    \eta=\frac{1}{2} - \frac{\epsilon}{4} - \frac{\epsilon^2}{8} + \frac{\epsilon^3}{8} + \frac{7\epsilon^4}{32}+....
\end{equation}
where the sign of $\epsilon$ is identical with the sign of $\Lambda$ (see eq. \eqref{epsdef}).

The determination of the magnitude of $ \eta\equiv \frac{R_{\text{vir}}}{R_{\text{max}}}$  and, consequently, the virial radius, $R_{p,\text{vir}}$, becomes feasible if $R_{p,\text{max}}$ is obtained along the lines discussed in the previous section using Equation \eqref{ratiorc} and the numerical determination of $\delta_{\text{max}}$. From Equation \eqref{rvirrmax} it becomes clear that when $\Lambda$, is positive, $\eta$ is less than $\frac{1}{2}$. Thus $R_{p,\text{vir}}$ is less than $\frac{1}{2}R_{p,\text{max}}$, meaning that to achieve virialization, mass shells are compelled to delve into deeper regions of the spherical overdensity to acquire the necessary velocity since the repulsive gravity of $\Lambda$ in this case needs stronger attractive gravity of matter to be balanced. Conversely, when $\Lambda$ is negative, it results in $\eta$ exceeding $\frac{1}{2}$. This indicates that $R_{p,\text{vir}}$ is larger than $\frac{1}{2}R_{p,\text{max}}$, meaning that mass shells secure the necessary velocity before reaching half of the turnaround radius $R_{p,\text{max}}$ since the gravitational force due to negative $\Lambda$ is attractive and amplifies the gravitational effects of matter.

The parameter $\epsilon$ can be obtained using Equation (\ref{eq009}) and the numerical determination of $\delta_{\text{max}}$ desscribed in the previous section as
\begin{equation}\label{rholrhom}
\epsilon(a_{\text{max}},\Omega_{\Lambda,0},\Omega_{m,0}) \equiv \frac{\rho_{\Lambda}}{\rho_{m,\text{max}}}=\frac{\rho_{\Lambda}}{\tilde{\rho}_{m,0}R^{-3}_{\text{max}}}=\frac{\omega}{1+\delta_{\text{max}}}a_{\text{max}}^3 
\end{equation}

We can now derive the density of virialized objects in terms of the cosmological parameters ${\Omega_{\Lambda,0}}$, ${\Omega_{m,0}}$ and the turnaround redshift $z_{\text{max}}=\frac{1}{a_{\text{max}}}-1$. Using Equations (\ref{ratiorc}), (\ref{rvirrmax}) and (\ref{rholrhom}) we may express the scale of the overdensity as
\be
R_{\text{vir}}\simeq\frac{(1+z_{\text{max}})^{-1}\left(\frac{M}{M_c}\right)^{1/3}}{\left[(1+\delta_{\text{max}})\Omega_{m,0}\right]^{1/3}}\left(\frac{1}{2}-\frac{1}{4}\frac{\omega}{1+\delta_{\text{max}}}(1+z_{\text{max}}^{-3})\right)
\ee
where we have kept terms up to first order in $\epsilon$ as obtained in Equation (\ref{rholrhom}).

\begin{figure}
\centering
\includegraphics[width = \columnwidth]{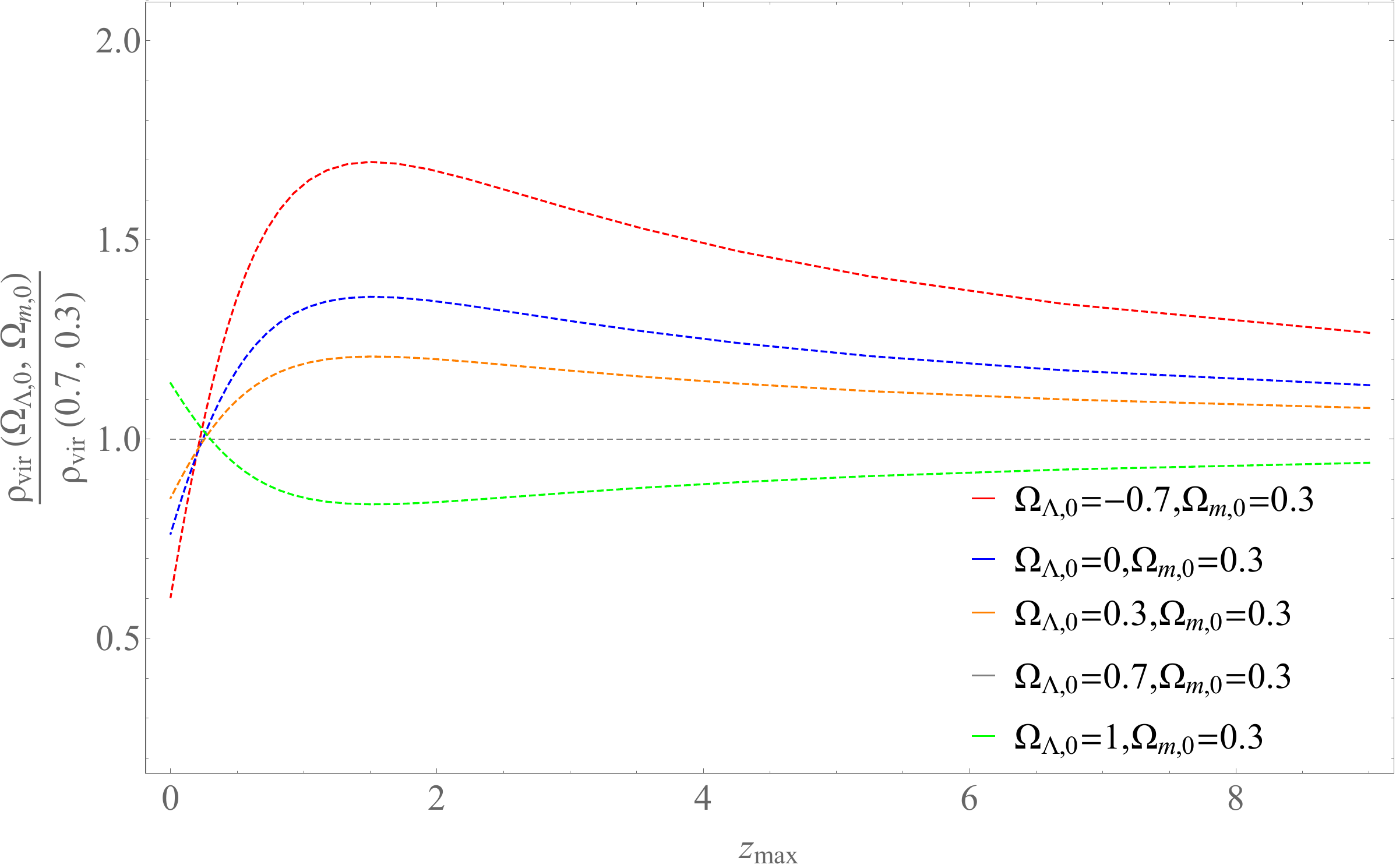}
\caption{The provided diagram offers a comparative representation of the ratio between virialized energy overdensity, given various numerical values for the density parameters $\Omega_{m,0}$, $\Omega_{\Lambda,0}$ and curvature $\Omega_{k}\neq 0$, and the virialized energy density within a $\Lambda$CDM model where $\Omega_{m,0}=0.3$ and $\Omega_{\Lambda,0}=0.7$. This comparison is illustrated with respect to the turn around redshift, denoted as $z_{\text{max}}$.}
\label{fig1}
\end{figure}

\begin{figure}
\centering
\includegraphics[width = \columnwidth]{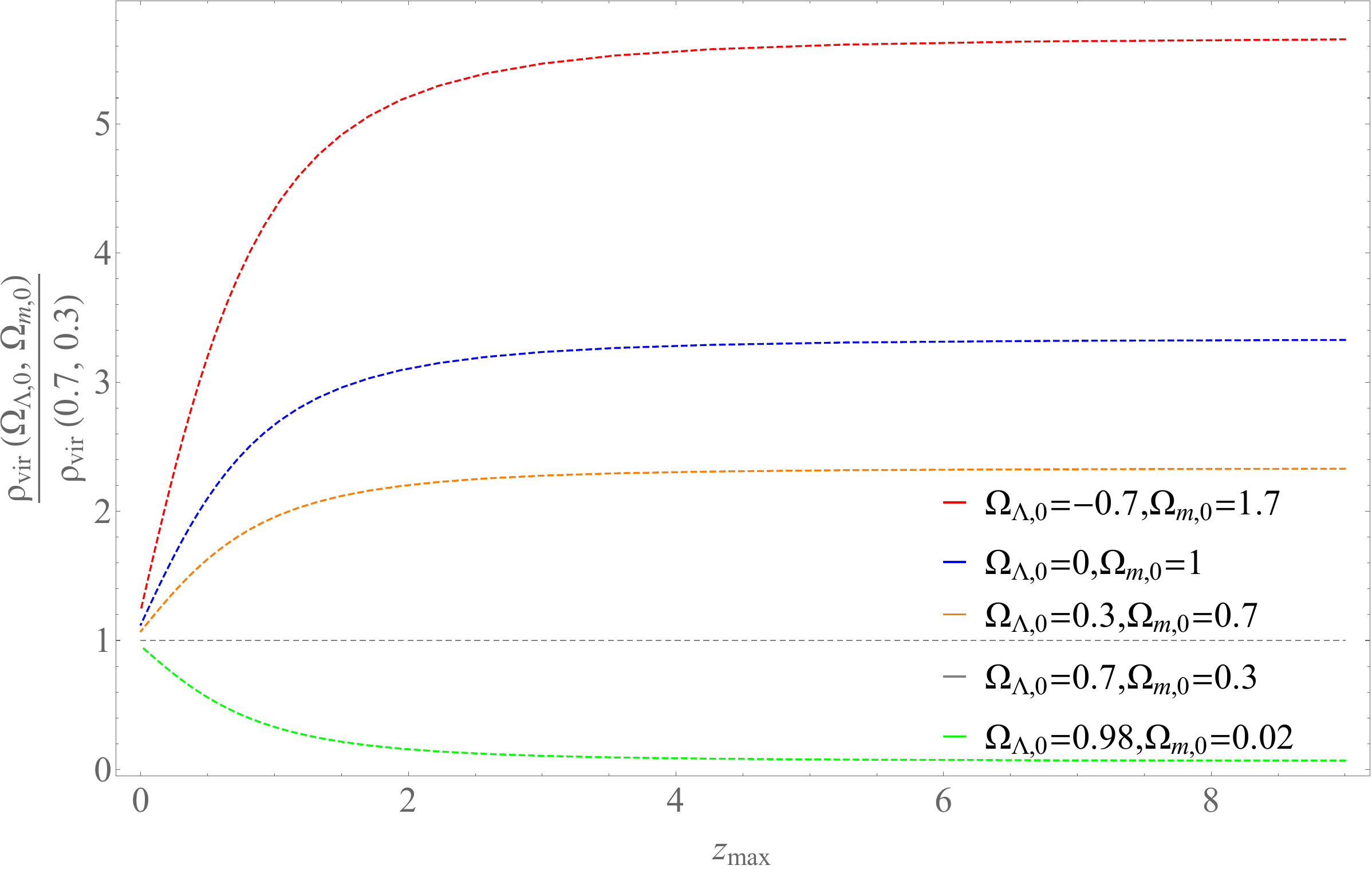}
\caption{The provided diagram offers a comparative representation of the ratio between virialized energy overdensity, given various numerical values for the density parameters $\Omega_{m,0}$, $\Omega_{\Lambda,0}$ and curvature $\Omega_{k}= 0$, and the virialized energy density within a $\Lambda$CDM model where $\Omega_{m,0}=0.3$ and $\Omega_{\Lambda,0}=0.7$. This comparison is illustrated with respect to the turn around redshift, denoted as $z_{\text{max}}$.}
\label{fig2}
\end{figure}

Since the density at virialization is $\rho_{\text{vir}}\sim R_{p,\text{vir}}^{-3}$ for a given overdensity mass, we can now find the predicted ratio of the virialized density in a given dark energy model over the corresponding density predicted by \plcdm . For example, for a general value of $\Lambda$, we have
\be
\frac{\rho_{\text{vir}}(z_{\text{max}},\Omega_{m,0},\Omega_{\Lambda,0})}{\rho_{\text{vir}}(z_{\text{max}},0.3,0.7)}=\left(\frac{R_{\text{vir}}(z_{\text{max}},0.3,0.7)}{R_{\text{vir}}(z_{\text{max}},\Omega_{m,0},\Omega_{\Lambda,0})}\right)^3
\label{rvirrat}
\ee
Thus we construct Fig. \ref{fig1} where we show this ratio of predicted virialized densities as a function of the turnaround redshift $z_{\text{max}}$, for fixed value of $\Omega_{m,0}=0.3$ and varying $\Omega_{\Lambda,0}$. Similarly, Fig. \ref{fig2} shows the same ratio varying both $\Omega_{m,0}$ and $\Omega_{\Lambda,0}$ in a flat universe $\Omega_{m,0} +\Omega_{\Lambda,0}=1$. Clearly, the reduction of the cosmological constant reduced the repulsive part of gravity leading to objects with higher virialized densities up to $80\%$ compared to \plcdm for $z_{\text{max}}\simeq 2$ for negative $\Omega_{\Lambda,0}=-0.7$.  This notable increase is suggesting a possibility for more prolific early galaxy formation under this class of models. This observation is in congruence with the recent revelations made by the James Webb Space Telescope \citep{Menci:2022wia,Biagetti:2022ode,Wang:2022jvx,Forconi:2023izg}.

\begin{figure}
\centering
\includegraphics[width = \columnwidth]{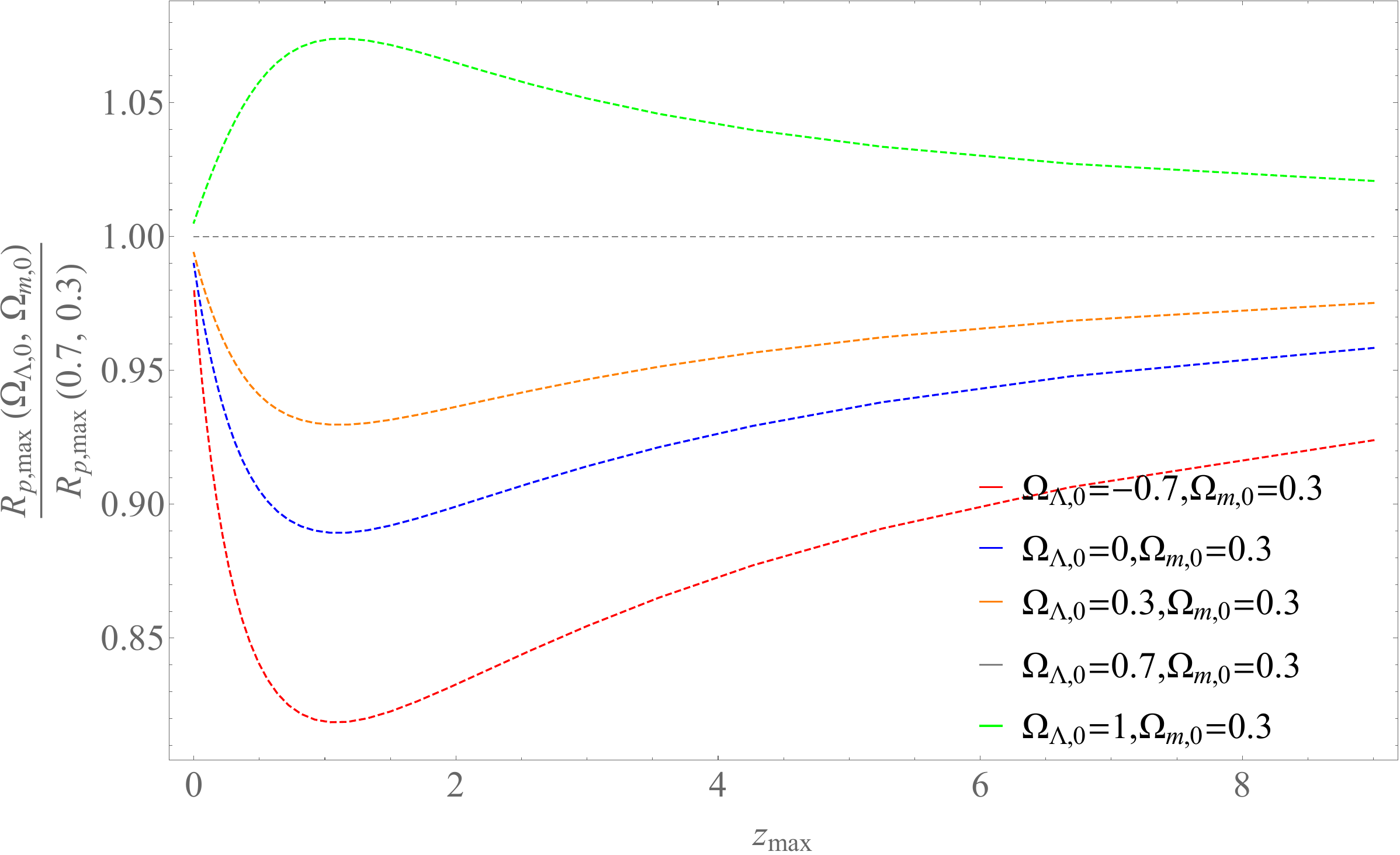}
\caption{The turnaround radius, evaluated over a range of density parameters $\Omega_{m,0}$ and $\Omega_{\Lambda,0}$, relative to the  turnaround radius within a \plcdm framework constituted by $\Omega_{m,0}=0.3$ and $\Omega_{\Lambda,0}=0.7$, in terms of the turnaround redshift $z_{\text{max}}$.}
\label{fig3}
\end{figure}

\begin{figure}
\centering
\includegraphics[width = \columnwidth]{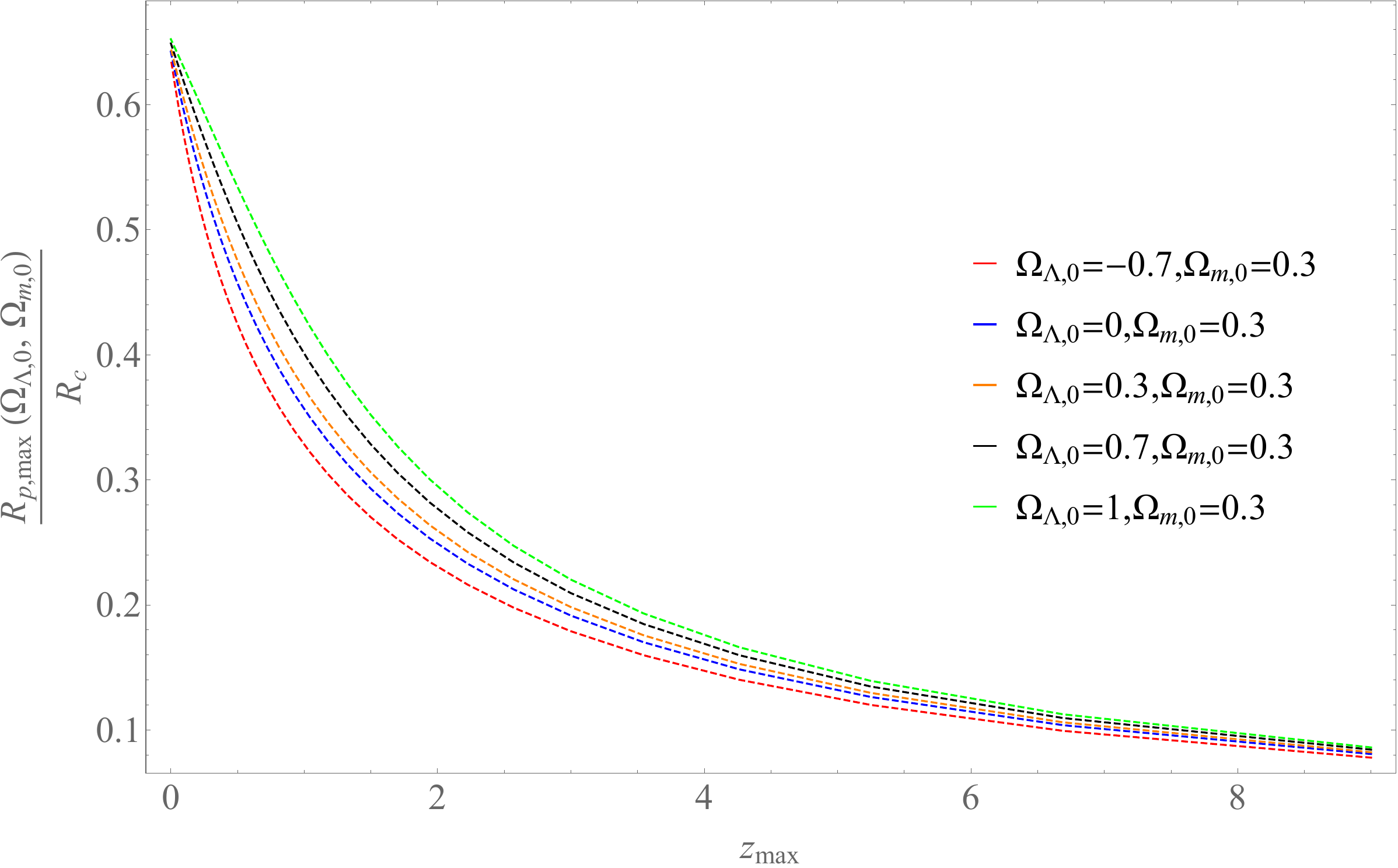}
\caption{The ratio between the turn around radius, considering a range of density parameters $\Omega_{m,0}$ and $\Omega_{\Lambda,0}$ relative to the length scale   $R_{c}\equiv\left(\frac{3 M_{c}}{4 \pi \tilde{\rho}_{\text{crit},0}}\right)^{\frac{1}{3}}$, in terms of the turnaround redshift $z_{\text{max}}$.}
\label{fig4}
\end{figure}

\begin{figure}
\centering
\includegraphics[width = \columnwidth]{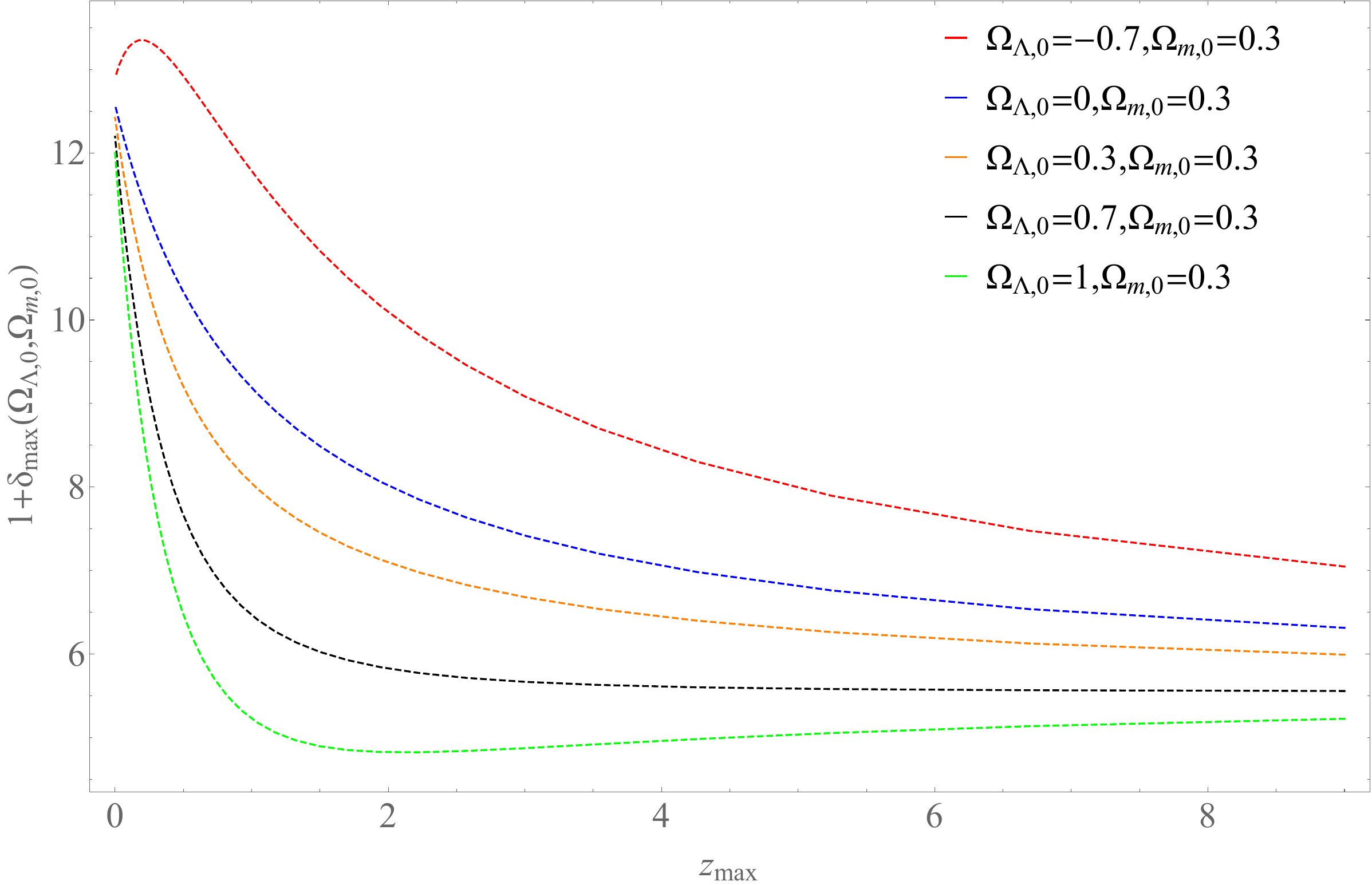}
\caption{The density contrast of $1+\delta_{\text{max}}(z_{\text{max}};\Omega_{\Lambda,0}, \Omega_{m,0})=\frac{\rho_{m,\text{max}}}{\tilde \rho_m}$ obtained by the numerical solution of Equation (\ref{drda6}).  This quantity is the energy overdensity at the time of turnaround ($t_{\text{max}}$) over the corresponding  matter energy density of the background at the turnaround time $t_{\text{max}}$, in terms of the turn around redshift $z_{\text{max}}$, for some values of the density parameters $\Omega_{m,0}$ and $\Omega_{\Lambda,0}$.}
\label{fig5}
\end{figure}

\begin{figure}
\centering
\includegraphics[width = \columnwidth]{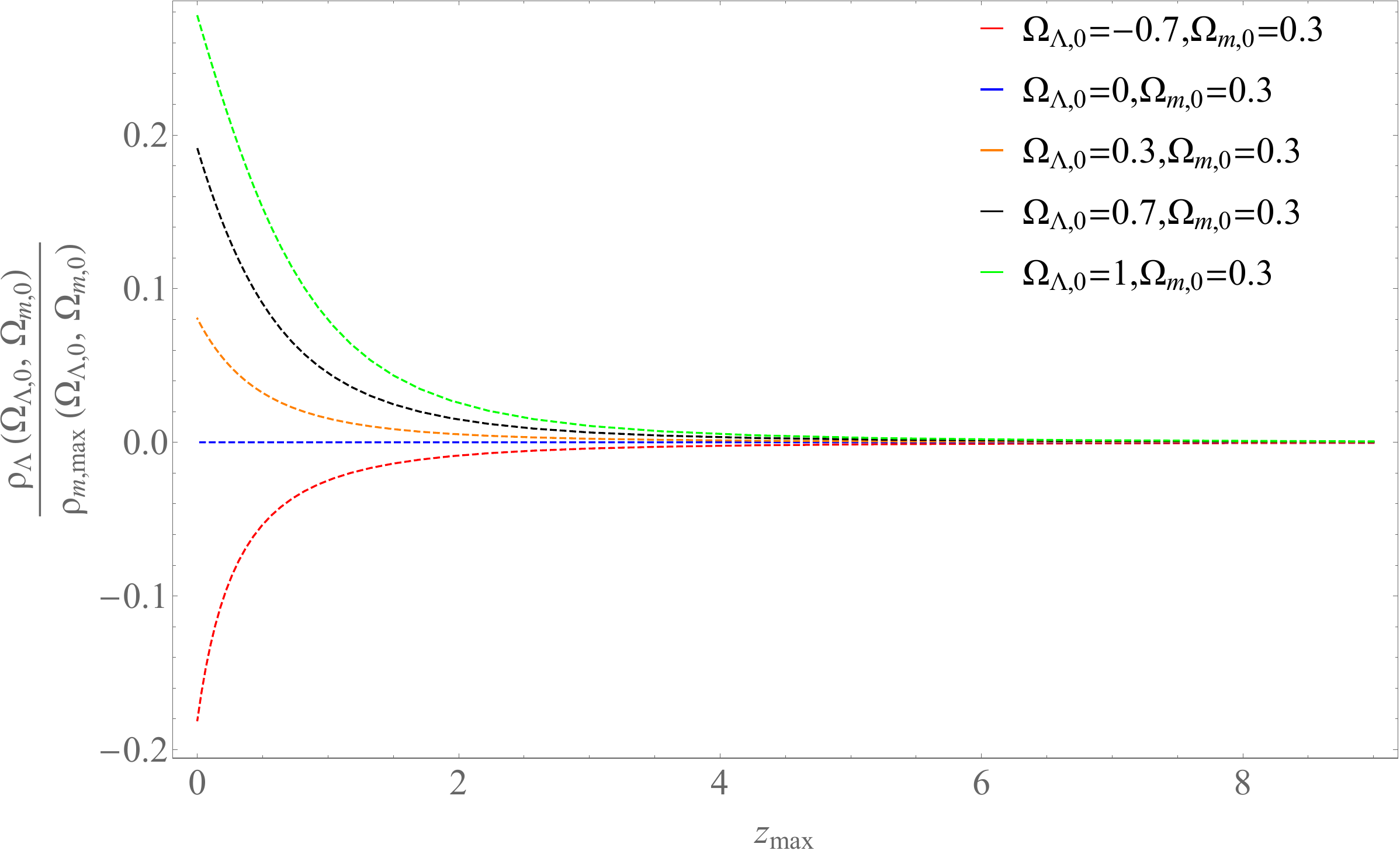}
\caption{The density contrast parameter $\epsilon (z_{\text{max}};\Omega_{\Lambda,0}, \Omega_{m,0})$, considering a range of density parameters $\Omega_{m,0}$ and $\Omega_{\Lambda,0}$, is shown through the dependency on the turnaround redshift $z_{\text{max}}$.}
\label{fig6}
\end{figure}

The corresponding turnaround radii ratios with respect to \lcdm are shown in Fig. \ref{fig3} where we obtain the physically anticipated result that the turnaround radius increases with $\Lambda$: In the presence of more repulsive gravity the shell expands more,  until the attractive part of gravity (matter) turns it around. The typical values of the turnaround radius $R_{\text{max}}$ in units of the typical cluster scale $R_c$ are shown in Fig. \ref{fig4} using eq. (\ref{ratiorc}) for $M=M_c$. The corresponding density ratios of the structures at turnaround with respect to the background matter density ($\frac{\rho_{m,\text{max}}}{\tilde \rho_m}=1+\delta_{\text{max}}$) obtained from the numerical solution of Equation \eqref{drda6}, are shown in Fig. \ref{fig5}. As expected, more repulsive gravity (higher $\Lambda$) leads to lower matter overdensities at turnaround. Finally in Fig. \ref{fig6} we show the dark energy density ratio with respect to the matter density inside the spherical overdensity as a function of the turnaround redshift $z_{\text{max}}$ as obtained from Equation (\ref{rholrhom}). As expected for $\Lambda>0$ the dark energy density is larger at late times and tends to dominate over the matter density at late turnaround times.

\section{Conclusion}

We have analyzed the impact of a variable, positive or negative cosmological constant on the nonlinear dynamics of structure formation. This class of models embedded in a sign switching cosmological constant model ($\Lambda_s\text{CDM}$),  has recently gained interest as a novel approach for the resolution of cosmological tensions \citep{Dutta:2019pio,Dutta:2018vmq,Adil:2023exv,Akarsu:2022typ} and for understanding galaxy formation and the large-scale structure of the Universe. We utilize the spherical collapse model and set up the framework for examining the effects of a negative $\Lambda$ on the collapse and subsequent virialized matter overdensities in the Universe. We find that a negative cosmological constant influences the behavior of the spherical overdensities, impacting their evolution and the process of virialization and we contrast it with \plcdm.  

For negative values of $\Lambda$ inspired by sign-switching models, which are proposed for resolving the Hubble and $S_8$ tensions, and for a preferred value of $\Omega_{\Lambda,0}=-0.7$ (with $\Omega_{m,0}$=0.3) \citep{Akarsu:2022typ}, we observe a significant amplification of the density of virialized clusters that can reach up to 80\% in comparison with the \plcdm model. This is evident for a turnaround redshift $z_{\text{max}} \gtrsim 2$, indicating a potential for more efficient early galaxy formation within this class of models. This aligns well with the recent discoveries made by the James Webb Space Telescope \citep{Menci:2022wia,Biagetti:2022ode,Wang:2022jvx,Forconi:2023izg}.

We have demonstrated that the observed turnaround and virialized radii and densities of collapsing clusters  constrain the value of the cosmological constant and dark energy properties independently from the standard cosmological distance probes. 

Through detailed analysis of the ratio of virialized energy overdensity for models characterized by varying values of the density parameters $\Omega_{m,0}$, $\Omega_{\Lambda,0}$, and curvature $\Omega_{k,0}$, we found that a decrease in the cosmological constant below its observed value in the standard \lcdm model results in a higher proportion of energy overdensity undergoing virialization.

These findings, albeit preliminary, suggest that a negative cosmological constant, $\Lambda$, could give rise to novel and distinguishable consequences with regard to the macroscopic structure of the Universe. 

Interesting extensions of the present analysis include the comparison with observed cluster and supercluster virialized and turnaround densities for the derivation of constraints on value of the cosmological constant. Other interesting extensions include the study of the virialization and turnaround densities of more general dark energy fluids including the virialization effect associated with the cosmological constant transitioning from negative to positive. Such a transition might indeed denote a discontinuity on the Hubble flow i.e a sudden cosmological singularity may occur \citep{Barrow:2004xh,Fernandez-Jambrina:2006tkb,Perivolaropoulos:2016nhp,Paraskevas:2023aae}, a phenomenon that could deform or dissociate some large-scale structures.

\section*{Acknowledgments}
We thank Özgür Akarsu and Ruchika, for their insightful discussions that inspired this project and influenced its direction. This research was supported by COST Action CA21136 - Addressing observational tensions in cosmology with systematics and fundamental physics (CosmoVerse), supported by COST (European Cooperation in Science and Technology). This project was also supported by the Hellenic Foundation for Research and Innovation (H.F.R.I.), under the "First call for H.F.R.I. Research Projects to support Faculty members and Researchers and the procurement of high-cost research equipment Grant" (Project Number: 789).

\section*{Data Availability Statement}
The Mathematica (v12) files used for the production of the figures and for derivation of the main results of the analysis can be found at \href{https://github.com/vaggelis01/Density-of-Virialized-Clusters}{this Github repository under the MIT license.} \\



\bibliographystyle{mnras}
\bibliography{main} 



\newpage
\appendix

\section{Examining the Overdensity at the Point of Turnaround within a Cosmological Constant-Driven Background}\label{AppA}

In accordance with the aforementioned notation and following the work of \cite{Tanoglidis:2016lrj}, the expression representing the homogeneous spherical overdensity, at the point of turnaround, can be written as 
\begin{equation}\label{A01}
\rho_{m}\left(a_{\text{max}}\right)=\tilde{\rho}_{m}\left(a_{\text{max}}\right)(1+\delta_{\text{max}})
\end{equation}
where $\tilde{\rho}_{m}$ denotes the energy density of the background spacetime, and $\rho_{m}$ represents the overdensity. Taking into account $a(t)$ as the scale factor of the background universe and $R(t)$ as the scale factor of the uniform spherical overdensity, and that equation (\ref{A01}) yields \citep{Pavlidou:2004vq,Tanoglidis:2016lrj}
\begin{equation}\label{B1}
\delta_{\text{max}}=\left(\frac{a_{\text{max}}}{R_{\text{max}}}\right)^3-1
\end{equation}
then, we can express matter energy densities $\rho_{m}=\tilde{\rho}_{m,0} R^{-3}$ and  $\tilde{\rho}_{m}=\tilde{\rho}_{m,0}a^{-3}$. Here, $\tilde{\rho}_{m,0}$ represents the present value of the energy density of the matter distribution. Additionally, the physical radius of the spherical overdensity is denoted as $R_{p}\equiv R_{p}(t)=\chi_{0}R(t)$, where, $R(t)$, the local scale factor and $\chi_{0}$ the corresponding comoving radius.

The dynamics of both $R(t)$ and $a(t)$ are determined by two Friedman equations which include spatial curvature matter density and cosmological constant. The Friedmann equation for the cosmological background is \citep{Eke:1996ds,Pavlidou:2004vq,Tanoglidis:2016lrj,Pavlidou:2020afx}
\begin{equation}\label{eqB1}
     \left(\frac{\dot{a}}{a}\right)^2=\frac{8\pi G}{3} \left(\tilde{\rho}_{m,0}a^{-3}+\rho_{\Lambda}+\frac{\tilde{\rho}_{\text{crit},0}-\tilde{\rho}_{m,0}-\rho_{\Lambda}}{a^2}\right)
\end{equation}
The energy density attributed to the cosmological constant is denoted by $\rho_{\Lambda}\equiv \frac{\Lambda}{8\pi G}$. Furthermore, the critical density of the background at the current epoch is \begin{equation}\tilde{\rho}_{\text{crit},0}\equiv \frac{3 H_{0}^2}{8\pi G}\end{equation}

We also introduce a term $\omega$, defined as the ratio of the energy density due to the cosmological constant and the present day background matter density, i.e. 

\begin{equation}
\omega\equiv \frac{\rho_{\Lambda}}{\tilde{\rho}_{m,0}}=\frac{\Omega_{\Lambda,0}}{\Omega_{m,0}}
\end{equation}
While the parameter associated with the spatial curvature of the cosmological background is further described as
\begin{equation}\xi\equiv\frac{\tilde{\rho}_{\text{crit},0}-\tilde{\rho}_{m,0}-\rho_{\Lambda}}{\tilde{\rho}_{m,0}}=\frac{1}{\Omega_{m,0}}-1-\omega 
\end{equation}
Considering the aforementioned details, the Friedman equation in Eq.~(\ref{eqB1}) is expressed as:
\begin{equation}\label{eqB3}
    \left(\frac{\dot{a}}{a}\right)^2=\frac{8\pi G}{3} \tilde{\rho}_{m,0}\left(a^{-3}+\omega+\frac{\xi}{a^2}\right)
\end{equation}

The corresponding Friedmann equation of the local scale factor $R(t)$, given the initial conditions, can be inferred as 
\begin{equation}\label{rescapp}
    \left(\frac{\dot{R}}{R}\right)^2=\frac{8\pi G}{3}\tilde{\rho}_{m,s} \left(R^{-3}+\frac{\rho_{\Lambda}}{\tilde{\rho}_{m,s}}-\frac{\tilde{\kappa}}{R^2}\right)
\end{equation}
Taking into account that, at the initial cosmic time $t_{s}$, the following equation  \begin{equation}\rho_{m,s} = \tilde{\rho}_{m,s} = \tilde{\rho}_{m,0}a_{s}^{-3}\end{equation}
 holds and upon rescaling with transformations \(a_{s} R \rightarrow R\) and \(\frac{\tilde{\kappa}}{a_{s}} \rightarrow \kappa\), we can express Equation ~(\ref{rescapp}) as:
\begin{equation}\label{eqB4}
    \left(\frac{\dot{R}}{R}\right)^2=\frac{8\pi G}{3}\tilde{\rho}_{m,0} \left(R^{-3}+\omega-\frac{\kappa}{R^2}\right)
\end{equation}
Consequently, by dividing \eqref{eqB3} by \eqref{eqB4}, we obtain the subsequent equation:
\begin{equation}\label{B8}
  \left(\frac{dR}{da}\right)^2 =\frac{R^2}{a^2}\frac{R^{-3}+\omega-\frac{\kappa}{R^2}}{a^{-3}+\omega+\frac{\xi}{a^2}}=\frac{a}{R}\frac{1+\omega R^3-\kappa R}{1+\omega a^3+\xi a}.
\end{equation}
Consider a turnaround point  where the derivative of the local scale factor with respect to the cosmological scale factor is \begin{equation}\frac{d R}{da}\bigg{|}_{a=a_{\text{max}}}=0\end{equation}
At this juncture, the radius attains a maximum value denoted as $R_{p,\text{max}}$, while the scale factor at that point is $R(a_{\text{max}})=R_{\text{max}}$. Under this condition, it becomes requisite to find a positive solution for the equation

\begin{equation}
1+\omega R_{\text{max}}^3-\kappa R_{\text{max}}=0
\end{equation}
which yields the value of the parameter $\kappa$ as a function of $\omega$ and $R_{\text{max}}$:

\begin{equation}\label{B10}
\kappa= \frac{\omega R^3_{\text{max}}+1}{R_{\text{max}}}.
\end{equation}
By choosing the positive square root of (\ref{B8}), we obtain
\begin{equation}\label{A014}
     \frac{dR}{da} =\sqrt{\frac{a}{R}\frac{1+\omega R^3-\kappa R}{1+\omega a^3+\xi a}}.
\end{equation}
Following the integration of equation (\ref{B8}) and by incorporating the initial condition $R(0)=0$, equation (\ref{B14}) is derived:

\begin{equation}\label{B14}
\int_{0}^{R}dR'\sqrt{\frac{R'}{1+\omega R'^3-\kappa R'}}=\int_{0}^{a}da'\sqrt{\frac{a'}{1+\omega a'^3+\xi a'}}.
\end{equation}
By invoking a substitution of variables where $u=\frac{R'}{R_{\text{max}}}$, and with $r$ defined as $r\equiv \frac{R}{R_{\text{max}}}$, applied to the left-hand side of equation (\ref{B14}), we derive the consequent expression delineated in equation (\ref{B15}):
\begin{equation}\label{B15}
\int_{0}^{R}dR'\sqrt{\frac{R'}{1+\omega R'^3-\kappa R'}}=\int_{0}^{r} du\sqrt{\frac{u}{\omega u^3- \kappa R_{\text{max}}^{-2}u+R_{\text{max}}^{-3}}}.
\end{equation}
Applying equation (\ref{B10}), with $\kappa R_{\text{max}}^{-2}=\omega+R_{\text{max}}^{-3}$, to equation (\ref{B15}) yields a simplified version of the integral. The entire simplification process is represented as follows:

\begin{equation*}
=\int_{0}^{r} du\sqrt{\frac{u}{\omega u^3- (\omega+R_{\text{max}}^{-3})u+R_{\text{max}}^{-3}}}
\end{equation*}
\begin{equation*}
=\int_{0}^{r} du\sqrt{\frac{u}{\omega u^{3}-\omega u-u R_{\text{max}}^{-3}+R_{\text{max}}^{-3}}}
\end{equation*}
\begin{equation*}
=\int_{0}^{r} du\sqrt{\frac{u}{\omega u(u-1)(u+1)- R_{\text{max}}^{-3}(u-1)}}
\end{equation*}
\begin{equation*}
=\int_{0}^{r} du\sqrt{\frac{1}{(1-u)}\frac{u}{R_{\text{max}}^{-3}-\omega u(u+1) }}
\end{equation*}

Further use of equation (\ref{B1}) and the realization that $R=R_{\text{max}}\implies r=1$ at the turnaround point, allows us to represent the integral as follows:

\begin{equation}\label{B16}
=\int_{0}^{1} du\sqrt{\frac{1}{\omega}\frac{1}{1-u}\frac{u}{\frac{1+\delta_{\text{max}}}{\omega a_{\text{max}}^{3}}-u(u+1)}}
\end{equation}
By implementing a substitution of variables with $y=\frac{a'}{a_{\text{max}}}$ and setting as $q\equiv \frac{a}{a_{\text{max}}}$ into the right-hand side of equation (\ref{B14}), the following expression (\ref{B17}) is derived:

\begin{equation}\label{B17}
\int_{0}^{a}da'\sqrt{\frac{a'}{1+\omega a'^3+\xi a'}}=\int_{0}^{q}dy \sqrt{\frac{y a^{3}_{\text{max}}}{\omega a^{3}_{\text{max}} y^{3}+\xi y a_{\text{max}}+1}}=
\end{equation}
\begin{equation*}
   = \int_{0}^{q}dy \sqrt{\frac{y }{\omega y^{3}+\xi y a_{\text{max}}^{-2}+a_{\text{max}}^{-3}}}=
\end{equation*}
By imposing the condition $a=a_{\text{max}}$, which consequently implies $q=1$, we ascertain the following:
\begin{equation*}
    =\int_{0}^{1}dy \sqrt{\frac{y }{\omega y^{3}+\xi y a_{\text{max}}^{-2}+a_{\text{max}}^{-3}}}
\end{equation*}
Therefore, by utilising Equation (\ref{B14}) in conjunction with Equations (\ref{B16}) and (\ref{B17}), the final integral expression (\ref{B18}) can be derived:

\begin{equation}\label{B18}
\int_{0}^{1} du\sqrt{\frac{1}{1-u}\frac{u}{\frac{1+\delta_{\text{max}}}{\omega a_{\text{max}}^{3}}-u(u+1)}} = \int_{0}^{1}dy \sqrt{\frac{\omega y }{\omega y^{3}+\xi y a_{\text{max}}^{-2}+a_{\text{max}}^{-3}}}
\end{equation}

At this point, through numerical integration, we are able to ascertain the value of overdensity $\delta_{\text{max}}$. Furthermore, if we were to integrate the Equation (\ref{A014}) for an arbitrary value of $R(a)<R_{\text{max}}$, then by proceeding with numerical integration until $q(a)=\frac{a}{a_{\text{max}}}$ and $r(a)=\frac{R(a)}{R_{\text{max}}}$ are reached, we could derive  the local scale factor $R\equiv R(a)$ in terms of the background scale factor $a$.
\begin{equation}\label{B019}
\int_{0}^{r(a)} du\sqrt{\frac{1}{1-u}\frac{u}{\frac{1+\delta_{\text{max}}}{\omega a_{\text{max}}^{3}}-u(u+1)}} = 
\end{equation}
\begin{equation*}
    =\int_{0}^{q(a)}dy \sqrt{\frac{\omega y }{\omega y^{3}+\xi y a_{\text{max}}^{-2}+a_{\text{max}}^{-3}}}
\end{equation*}

\section{Virialized Dark Matter within a Homogeneous Dark Energy-Driven Background}\label{AppB}

Let us consider $\Phi$ to be the gravitational potential. The extraction of the Newtonian limit from the Einstein field equations yields a corresponding Poisson equation which fundamentally encapsulates pressure as a source of a gravitational field as

\begin{equation}
\nabla^2 \Phi = 4\pi G \left(\rho + 3\frac{p}{c^2}\right)
\end{equation}
We proceed by incorporating the equation of state, formulated as $p=w \rho c^2$, into our discussion. This inclusion allows for a reworking of the original Poisson equation into the form:

\begin{equation}
\nabla^2 \Phi = 4\pi G \rho(1+3 w)
\end{equation}
Assuming a theoretical framework wherein the \textit{homogeneous energy density is distributed spherically within a radius} $R$ and if the gravitational potential is $\Phi\equiv\Phi(r)$, then the solution to the Poisson equation manifests as follows, for $r\geq R$

\begin{equation}
    \Phi_{>}(r)=-\frac{C_{>}}{r}+D_{>}
\end{equation}
The condition $\Phi(r\to \infty)=0$ necessitates the setting of the constant $D_{>}$ to be identically zero. For the case of $r\leq R$, the solution can be formulated as:

\begin{equation}
\Phi_{<}(r)=2 \pi G \rho (1+3w)\frac{r^2}{3}-\frac{C_{<}}{r}+D_{<}
\end{equation}
The constraints $\Phi(r=0)<\infty$ and $\Phi_{>}(R)=\Phi_{<}(R)$ necessitate $C_{<}=0$ and lead to the following equation, respectively:

\begin{equation}\label{eq5}
2 \pi G \rho (1+3w)\frac{R^2}{3}+D_{<}=-\frac{C_{>}}{R}
\end{equation}
Moreover, the continuity of $\frac{d\Phi_{>}}{dr}\big{|}_{r=R}=\frac{d\Phi_{<}}{dr}\big{|}_{r=R}$, gives the following relation:

\begin{equation}\label{eq6}
4\pi G \rho (1+3w)\frac{R}{3}=\frac{C_{>}}{R^2}
\end{equation}
By incorporating equations (\ref{eq5}) and (\ref{eq6}), we arrive at the following expressions:

\begin{equation}
C_{>}=4\pi G \rho (1+3w)\frac{R^3}{3}
\end{equation}

\begin{equation}
D_{<}=-2\pi G \rho (1+3w)R^2
\end{equation}
Consequently, considering the prescribed boundary conditions, the resolution of the Poisson equation is given by:

\begin{equation}\label{eqA9}
\Phi(r)=\begin{cases}
    -2 \pi G \rho (1+3w)(R^2-\frac{r^2}{3})\quad r\leq R\\
-4\pi G \rho (1+3w)\frac{R^3}{3r}\quad r\geq R
    \end{cases}
\end{equation}
The gravitational potential $\Phi_{\text{DM}}$ of the dark matter is calculated by imposing $w=0$ at the equation (\ref{eqA9})

\begin{equation}
\Phi_{\text{DM}}(r)=\begin{cases}
    -2 \pi G \rho_{\text{DM}} (R^2-\frac{r^2}{3})\quad r\leq R\\
-4\pi G \rho_{\text{DM}} \frac{R^3}{3r}=-\frac{GM}{r}\quad r\geq R
    \end{cases}
\end{equation}
Here, the term $\rho_{\text{DM}}$ refers to the density of dark matter and the corresponding gravitational potential is denoted by $\Phi_{\text{DM}}$.

Moreover, in the case of \textit{homogeneity in the energy density of dark energy throughout the universe}, it becomes necessary to establish the appropriate boundary conditions i.e. $\Phi(r=0)<\infty$ and $\Phi(r= 0)=0$ in the solution \begin{equation}\Phi(r)=2 \pi G \rho (1+3w)\frac{r^2}{3}-\frac{C}{r}+D\end{equation}
which they imply $C=0$ and $D=0$. The gravitational potential $\Phi_{\text{DE}}$ for homogeneous dark energy is calculated as
\begin{equation}
   \Phi_{\text{DE}}= 2 \pi G \rho_{\text{DE}} (1+3w)\frac{r^2}{3}
\end{equation}
Here, $\rho_{\text{DE}}$ represents the density of dark energy and $\Phi_{\text{DE}}$ denote the corresponding gravitational potential. Drawing from the works of \cite{Lahav:1991wc,Iliev:2001he,Maor:2005hq,Saha:2023zos}, we describe the gravitational potential of a dual-component system consisting of dark energy and dark matter, where only the dark matter component undergoes virialization i.e the homogeneity of dark energy density, which seamlessly permeates from the external surroundings into the over-dense regions, indicates that dark energy does not coalesce or virialize with dark matter. The gravitational potential energy is

\begin{equation}\label{eqA12}
U_{\text{system}}=\frac{1}{2}\int_{V} \rho_{\text{DM}} \Phi_{\text{DM}} \, dV+\int_{V} \rho_{\text{DM}} \Phi_{\text{DE}} \, dV
\end{equation}
We have ruled out the possibility that dark energy can simultaneously cluster with dark matter and reach a virialized state. Notably, the interaction term lacks the coefficient \( \frac{1}{2} \). This omission can be ascribed to the distinctive character of the interaction between the different entities: dark matter and dark energy. Consequently, there's no need for the factor that usually compensates for double counting in identical pairwise interactions, as seen in the first term of equation \eqref{eqA12}.

 Proceeding under the premise that Equation (\ref{eqA12}) is applicable, we carry on with the subsequent analysis. For a homogeneous spherical dark matter distribution of a radius $R_p$ and mass $M$, the energy density is \begin{equation}\rho_{\text{DM}}=\frac{M}{\frac{4}{3}\pi R_{p}^{3}}\end{equation} and the first term is evaluated as

\begin{equation}
    \frac{1}{2}\int_{V} \rho_{\text{DM}}\Phi_{\text{DM}} \, dV=- 4\pi^2 G  \bigg(\frac{M}{\frac{4}{3}\pi R_{p}^3}\bigg)^2\int_{0}^{R_{p}} \left(R_{p}^2-\frac{r^2}{3}\right)r^2 \,dr=
\end{equation}
\begin{equation*}
    =-\frac{3}{5}\frac{GM^2}{R_{p}}
\end{equation*}
If $a(t)$ is the scale factor of the background universe and $R_{p}(t)$ the physical radius of the spherical symmetric homogeneous distribution of dark matter, then the energy densities, relative to their respective values at the point of turnaround, can be expressed as follows:

\begin{equation}\label{eqA14}
\frac{\rho_{\text{DE}}(t)}{\rho_{\text{DE}}(t_{\text{max}})}=\left(\frac{a(t)}{a(t_{\text{max}})}\right)^{-3(1+w)}
\end{equation}
 
\begin{equation}\label{eqA15}
\frac{\rho_{\text{DM}}(t)}{\rho_{\text{DM}}(t_{\text{max}})}=\left(\frac{R_{p}(t_{\text{max}})}{R_{p}(t)}\right)^3
\end{equation}
Assuming $\epsilon$ to be the ratio of the energy densities of dark energy to dark matter at the point of turnaround, i.e. \begin{equation}\epsilon\equiv\frac{\rho_{\text{DE}}(t_{\text{max}})}{\rho_{\text{DM}}(t_{\text{max}})}\end{equation}
we can rewrite the ratio of the energy densities at any time $t$ by using equations (\ref{eqA14}),(\ref{eqA15}) as:
\begin{equation}
\frac{\rho_{\text{DE}}(t)}{\rho_{\text{DM}}(t)}=\epsilon\left(\frac{a(t)}{a(t_{\text{max}})}\right)^{-3(1+w)}\left(\frac{R_{p}(t)}{R_{p}(t_{\text{max}})}\right)^3
\end{equation}

If we denote as $a(t_{\text{max}})\equiv a_{\text{max}}$ and $R_{p,\text{max}}\equiv R_{p}(t_{\text{max}})$, then the second term, involving the integral over the volume of $\rho_{\text{DM}}\Phi_{\text{DE}}$, can then be written as: 
\begin{equation}
\int_{V} \rho_{\text{DM}} \Phi_{\text{DE}}\, dV=\frac{2 \pi}{5} G \frac{\rho_{\text{DE}}}{\rho_{\text{DM}}} (1+3w)R_{p}^2 =
\end{equation} 
\begin{equation*}
    =-\frac{3}{5}\frac{GM^2}{R_{p}}\bigg{[}-\frac{1}{2} (1+3w)\epsilon \left(\frac{a}{a_{\text{max}}}\right)^{-3(1+w)}\left(\frac{R_{p}}{R_{p,\text{max}}}\right)^3\bigg{]}
\end{equation*}
The potential energy of the system is 
\begin{equation}
    U_{\text{system}}=-\frac{3}{5}\frac{GM^2}{R_{p}}\bigg{[}1+\theta \left(\frac{a}{a_{\text{max}}}\right)^{-3(1+w)}\left(\frac{R_{p}}{R_{p,\text{max}}}\right)^3\bigg{]}
\end{equation} where we have denoted as \begin{equation}\theta\equiv-\frac{1}{2}(1+3w)\epsilon\end{equation} 
The conversation of energy of the system $E$ gives that \begin{equation}E=(K_{\text{system}}+U_{\text{system}})|_{R_{p}=R_{p,\text{vir}}}=U_{\text{system}}|_{R_{p}=R_{p,\text{max}}}\end{equation}
Concurrently, according to the Virial theorem, the kinetic energy $K$ is defined as $K=\frac{R_{p}}{2}\frac{dU}{dR_{p}}$. These two premises collectively imply 
\begin{equation}\label{eq17}
    \bigg(\frac{R}{2}\frac{dU_{\text{system}}}{dR}+U_{\text{system}}\bigg)\bigg|_{R_{p}=R_{p,\text{vir}}}=U_{\text{system}}\big{|}_{R_{p}=R_{p,\text{max}}}
\end{equation}

The gravitational potential energy of the system at the moment of $t_{\text{max}}$ is

\begin{equation}
    U_{\text{system}}\big{|}_{R_{p}=R_{p,\text{max}}}=-\frac{3}{5}\frac{GM^2}{R_{p,\text{max}}}(1+\theta )
\end{equation}
and at the moment of virialization
\begin{equation}
   U_{\text{system}}\big|_{R_{p}=R_{p,\text{vir}}}= -\frac{3}{5}\frac{GM^2}{R_{p,\text{vir}}}\left( 1+\theta \eta^3 \left(\frac{a_{\text{vir}}}{a_{\text{max}}}\right)^{-3(1+w)}\right)
\end{equation}
while
\begin{equation}
    \frac{R}{2}\frac{dU_{\text{system}}}{dR}\bigg{|}_{R=R_{p,\text{vir}}}=-\frac{3 G M^2}{5 R_{p,\text{vir}}}\left( \theta \eta^3 \left(\frac{a_{\text{vir}}}{a_{\text{max}}}\right)^{-3(1+w)} -\frac{1}{2} \right)
\end{equation}
and the term on the left side of the equation (\ref{eq17}) yields
\begin{equation}
   \bigg(\frac{R_{p}}{2}\frac{dU_{\text{system}}}{dR_{p}}+U_{\text{system}}\bigg)\bigg|_{R_{p}=R_{p,\text{vir}}}= 
\end{equation}
\begin{equation*}
     =-\frac{3 G M^2}{10 R_{p,\text{vir}}}\left(1 +4\theta \left(\frac{a_{\text{vir}}}{a_{\text{max}}}\right)^{-3(1+w)} \left(\frac{R_{p,\text{vir}}}{ R_{p,\text{max}}}\right)^3\right)
\end{equation*}
By denoting as \begin{equation}\eta\equiv \frac{R_{p,\text{vir}}}{R_{p,\text{max}}}\end{equation} then the equation (\ref{eq17}) gives the following \cite{Lahav:1991wc,Iliev:2001he,Maor:2005hq,Saha:2023zos}
\begin{equation}\label{B24}
    4 \theta \eta^3 \left(\frac{a_{\text{vir}}}{a_{\text{max}}}\right)^{-3(1+w)}-2 (1+\theta)\eta +1=0
\end{equation}

\begin{figure}
\centering
\includegraphics[width = \columnwidth]{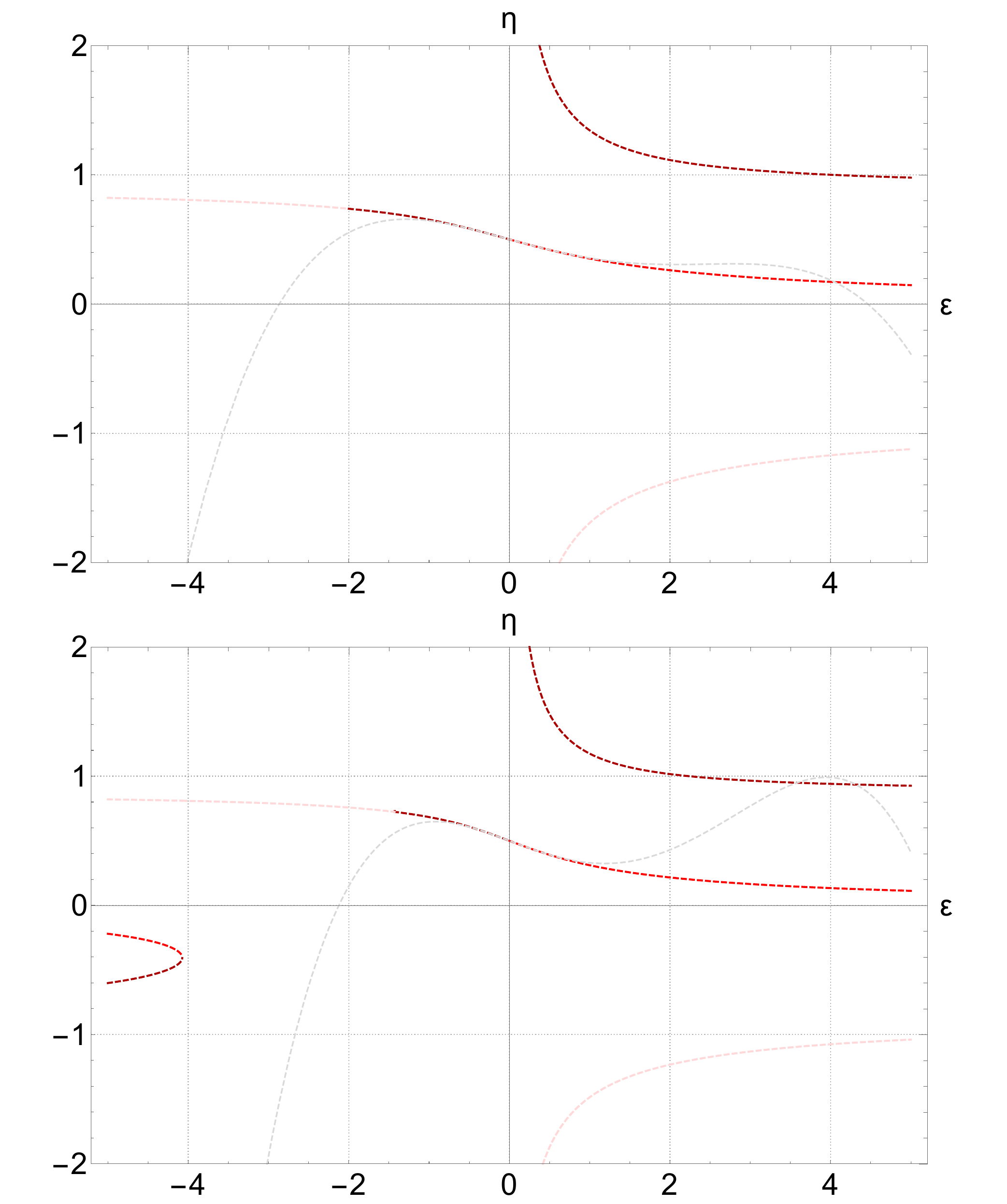}
\caption{This figure showcases three distinct solutions (shades of red) derived from equation (\ref{B24}). The \textit{upper} figure assumes parameters $w=-\frac{2}{3},a_{\text{vir}}=0.8, a_{\text{max}}=0.5$, and the \textit{lower} figure assumes $w=-0.8,a_{\text{vir}}=0.4, a_{\text{max}}=0.2$ for illustrative purposes. A \textit{light gray} line, representing a fourth order approximation from equation (\ref{eqA19}), is also evident in both graphs, further validating the approximated solution of (\ref{eqA19}) for $\epsilon<<1$.}
\label{figure:solutions}
\end{figure}

Given the inherent intricacy of the solution, the customary approach entails using a series expansion in terms of $\theta$. By designating as \begin{equation}\alpha\equiv\left(\frac{a_{\text{vir}}}{a_{\text{max}}}\right)^{-3(1+w)}\end{equation}
then, for both positive and negative values of $\theta$, Equation (\ref{B24}) delivers the following result:

\begin{equation}\label{eqA19}
   \eta=\frac{1}{2} + \frac{1}{4} \theta \left(-2 + \alpha\right) + \frac{1}{8} \theta^2 \left(-2 + \alpha\right) \left(-2 + 3 \alpha\right) + 
\end{equation}
\begin{equation*}
    +\frac{1}{8} \theta^3 \left(-2 + \alpha\right) \left(2 - 9 \alpha + 6 \alpha^2\right) + 
\end{equation*}

\begin{equation*}
    +\frac{1}{32} \theta^4 \left(-2 + \alpha\right) \{-8 + \alpha [76 + 5 \alpha \left(-26 + 11 \alpha\right)]\}+...
\end{equation*}

 In the specific case of a cosmological constant where $w=-1$, it logically infers that $\theta=\epsilon$, and consequently, $\alpha=1$. Moreover, we can interpret a condition where $\epsilon>0$ as being equivalent with a situation where the cosmological constant is positive ($\Lambda>0$). Conversely, if $\epsilon<0$, it equates to a scenario characterized by a negative cosmological constant ($\Lambda<0$). The equation (\ref{eqA19}) subsequently resulting in
\begin{equation}\label{eqA20}
    \eta=\frac{1}{2} - \frac{\epsilon}{4} - \frac{\epsilon^2}{8} + \frac{\epsilon^3}{8} + \frac{7\epsilon^4}{32}+....
\end{equation}
\begin{figure}
\centering
\includegraphics[width = \columnwidth]{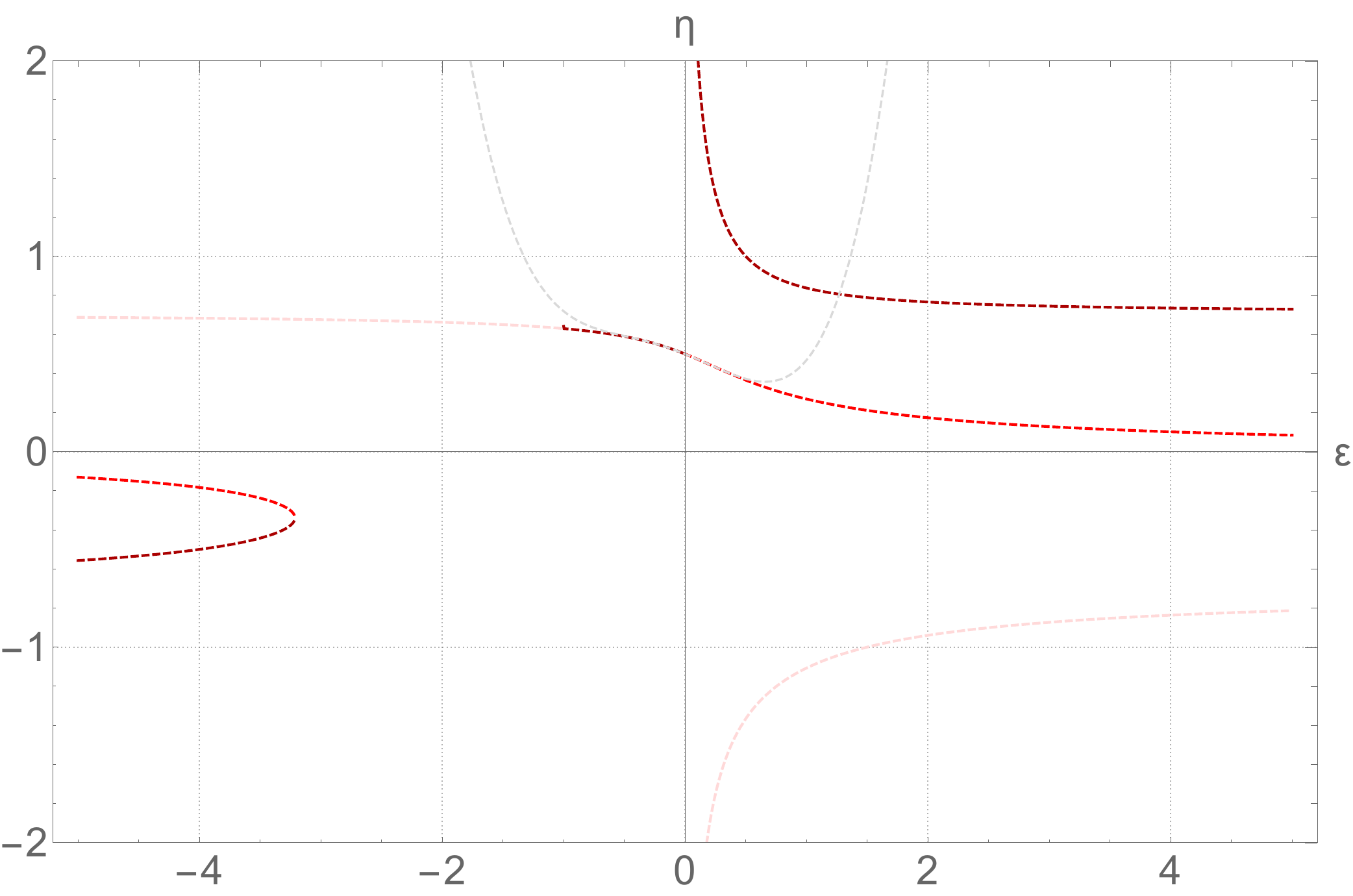}
\caption{Three solutions (shades of red) derived from equation (\ref{B24}), by assuming a value of $w=-1$.  Additionally, a line rendered in \textit{light gray} within the plot delineates an approximation up to the fourth order, as outlined in equation (\ref{eqA20}). }
\label{figure:solutions1}
\end{figure}

\bsp	
\label{lastpage}
\end{document}